\documentclass[11pt,a4paper]{article} 
\pdfoutput=1
\usepackage[no-natbib-sort]{my-jheppub}

\usepackage{amsmath, amssymb}
\usepackage{mathpazo}
\usepackage{environ}
\usepackage{mathrsfs}
\usepackage{array,arydshln}

\usepackage{graphicx,epsfig}
\usepackage{epic}
\usepackage{youngtab}
\usepackage{float}
\usepackage{color}
\definecolor{maroon}{rgb}{0.8,0.3,0.}

\usepackage{slashed}

\usepackage[nodayofweek]{date time}

\usepackage{hyperref}

\usepackage{aurical}
\usepackage[T1]{fontenc}

\newcommand{\be}{\begin{equation}}
\newcommand{\ee}{\end{equation}}

\newcommand{\ads}{AdS$_5\times S^5$\ }

\newcommand{\mc}{\mathcal }
\newcommand{\Z}{\mathcal{Z}}
\newcommand{\N}{\mathcal{N}}
\newcommand{\E}{{\mathcal E}}

\newcommand{\s}{{\rm s}}

\newcommand{\bv}{\begin{verbatim}}
\newcommand{\ev}{\end{verbatim}}

\newcommand{\red}[1]{\textcolor{red}{#1}}

\def \del{ \partial}
\def \la {\label}
\newcommand{\rf}[1]{(\ref{#1})}
\def\ov{\over}
\def\no{\nonumber} \def \aa {{\rm a}}
\def \ci {\cite}
\def \p {\phi}
\def \m {\mu}\def \n {\nu} 
\def \ed {\end{document}}
\def \l {\lambda} \def \r {\rho} 

\def \foot {\footnote}
 
\def \dd {{\rm d}}

\def \D {\Delta} 
\def \vp {\varphi} 
 \def \ha {{{1 \ov 2}}}

\def \cc {{\rm c}}
\def \aa {{\rm a} }

\def \De {\Delta} 
\def \ads {AdS$_{5}$\ }
\def \te {\textstyle} \def \iffa {\iffalse} 
\def \ha {{\te {1 \ov 2}}}
\def  \ba { \begin{align} }
\def  \ea { \end{align} }

\def \cc    {{\rm c}} 
\def \aa  {{\rm a}}
\def \kk  {{\rm k}}

\def \k {\kappa} \def \r {\rho}

\def \edd {\end{document}} 
\def \td {\tilde}

\def \KK {{\rm K}}

\def \iffa  {\iffalse}

\title{Conformal a-anomaly of some   non-unitary \\  6d  superconformal  theories }
\author[a]{Matteo Beccaria} 
\author[b]{, Arkady A. Tseytlin\footnote{Also at Lebedev Institute, Moscow}} 

\abstract{ 
We compute  the  conformal   anomaly  a-coefficient  for some 
 non-unitary (higher derivative or non-gauge-invariant) 6d conformal fields
 and their  supermultiplets. We use the  method 
 based on a connection between 6d determinants on S$^6$ and 
 7d determinants on AdS$_7$. We find,   in  particular,  that   (1,0)   supermultiplet  containing 4-derivative   gauge-invariant conformal vector  has precisely the value  of a-anomaly 
 as attributed in   \href{http://arxiv.org/abs/1506.03807}{arXiv:1506.03807} 
  (on the basis of R-symmetry and gravitational  't Hooft  anomaly matching) to the   standard (1,0) vector multiplet. 
 We also  show  that      higher  derivative (2,0)  6d conformal supergravity   coupled 
 to exactly 26   (2,0)   tensor multiplets   has vanishing  a-anomaly  (and also vanishing Casimir energy on 5-sphere).
 This is the  6d counterpart of the   known  fact of  cancellation of the conformal anomaly in the  4d  system of   $\N=4$   conformal supergravity   coupled to 4   vector $\N=4$  multiplets.    In  the case when 5 of  tensor multiplets  are chosen to be  ghost-like  and the conformal  symmetry is spontaneously broken  by a  quadratic  scalar constraint  the resulting IR  theory  may be identified  with  (2,0) 
  Poincar\'e  supergravity  coupled to 21=26-5  tensor multiplets. The latter  theory is known to be special -- it is   gravitational anomaly free  and results upon compactification of  10d type IIB   supergravity on K3.
  \vfill }
 
 \iffa 
 We compute   conformal   anomaly  a-coefficient  for some 
 non-unitary (higher derivative or non-gauge-invariant) 6d conformal fields
 and their  supermultiplets. We use the  method 
 based on a connection between 6d determinants on S^6 and 
 7d determinants on AdS_7. We find,   in  particular,  that (1,0)   supermultiplet  containing 4-derivative   gauge-invariant conformal vector  has precisely the value  of a-anomaly 
 as attributed in  arXiv:1506.03807
  (on the basis of R-symmetry and gravitational  't Hooft matching) to the   standard (1,0) vector multiplet. 
 We also  show  that      higher  derivative (2,0)  6d conformal supergravity   coupled 
 to exactly 26   (2,0)   tensor multiplets   has vanishing  a-anomaly  (and also vanishing Casimir energy on 5-sphere). This is the  6d counterpart of the   known  fact of  cancellation of the conformal anomaly in the  4d  system of    N=4   conformal supergravity   coupled to 4   vector N=4  multiplets.    In  the case when 5 of  tensor multiplets  are chosen to be  ghost-like  and the conformal  symmetry is spontaneously broken  by a  quadratic  scalar constraint  the resulting IR  theory  may be identified  with  (2,0)  Poincare  supergravity  coupled to 21=26-5  tensor multiplets. The latter  theory is known to be special -- it is   gravitational anomaly free  and results upon compactification of  10d type IIB   supergravity on K3. 
\fi

\affiliation[a]{Dipartimento di Matematica e Fisica Ennio De Giorgi,\\
Universit\`a del Salento \& INFN, Via Arnesano, 73100 Lecce, 
Italy} 

\affiliation[b]{The Blackett Laboratory, Imperial College, London SW7 2AZ, U.K.}

\emailAdd{matteo.beccaria@le.infn.it}
\emailAdd{tseytlin@imperial.ac.uk}

\allowdisplaybreaks

\begin{document}



 \begin{flushright}\small{Imperial-TP-AT-2015-{03}}\end{flushright}				

\maketitle

\flushbottom


\newcommand{\hh}{{\textstyle\frac{1}{2}}}

\def \adss {$AdS_5 \times S^5$ }
\def \adsss {$AdS_7 \times S^4$ }
\def \adse {$AdS_7$ }
\def \ads {$AdS_5$ }
\def \adt {$AdS_3$ }
\def \bh {{\bar h}} \def \bfh {{\bf{h}}}
\def \G {\Gamma}

\section{Introduction}

There has been some  recent interest  in   conformal   a-anomalies  of 
supersymmetric CFTs  in $d=6$ (see,  {\em e.g.},  \cite{Cordova:2015vwa,Cordova:2015fha,HH}  and refs. therein). 
This   motivates revisiting the question about  the computation of conformal anomalies 
for different types of  free 6d conformal fields. 
We shall relax the condition of unitarity  as   higher-derivative non-unitary   CFTs  may  be of  
interest, {\em e.g.}, 
as formal UV completions of some 
low-energy  
models  or 
as    6d counterparts of  some 
(higher spin)  theories in AdS$_7$.

Given a  conformal field  that can be coupled  to the metric in a reparametrization-invariant way, its
action on a curved background should be Weyl invariant. 
The conformal anomaly of a classically Weyl invariant theory in 6d has the following general form
 \cite{Bonora:1985cq,Deser:1993yx,Bastianelli:2000hi}  
\be
\label{1.1}
(4\pi)^{3} \langle T \rangle = b_6= \aa\,\E_{6}+ W_6+ 
D_6   \ , \ \ \ \ \ \  \ \ \  \ \ \ \ \ \    W_6=\cc_{1}\,I_{1}+\cc_{2}\,I_{2}+\cc_{3}\,I_{3}\ .
\ee
Here, $\E_{6} = -\epsilon_{6}\epsilon_{6}RRR$ is the Euler density in six dimensions, 
$W_6$   is a combination of   three independent   Weyl  invariants   built out  of  the Weyl tensor
 ($I_3\sim C \nabla^2 C, \ I_1,I_2 \sim CCC$)
and $D_6$ is a total derivative  term  (which is scheme-dependent), see  \cite{Bastianelli:2000hi} for details. 

Conformal anomalies for the simplest  6d free conformal 
fields (scalar, spinor and 2nd rank antisymmetric tensor)   were   found  in \cite{Bastianelli:2000hi}.  In particular,
 for the  (2,0) tensor multiplet\foot{We shall follow the notation of \ci{Bastianelli:2000hi} in which  
 a-anomaly  of a unitary  scalar is  negative in $d=6$ as opposed 
 to the  standard choice  of a>0 in $d=4$.}
 \be 
(2,0): \ \ \  W_6=  \cc { \cal W}_6 \ , \ \qquad    {\cal W}_6 =   96 I_1  + 24 I_2 - 8 I_3 \ , 
 \ \ \  \ \ \quad \cc= - { 1 \ov 288} \ , \ \ \ \   \aa=-{7\ov 1152} \ . \la{444} 
 \ee
   Let us note that while in the  case of (2,0) supersymmetry there  should be  a single 
superinvariant   containing   $W_6$ combination, {\em i.e.}  there  should  be 
 a single c-coefficient,  in  the  (1,0)  case there   should be  apparently two.
As follows from the results 
of \cite{Bastianelli:2000hi}, 
for  conformal anomalies   of both tensor and scalar (1,0) multiplets   the coefficients 
$\cc_i$ satisfy 
$\cc_1=2 \cc_2 - 6 \cc_3$.
 
The   a-coefficient  for  conformal higher-spin symmetric tensor   gauge fields 
was computed  in \cite{Tseytlin:2013fca},  both directly (from the partition function on $S^6$)   and  also using the 
 AdS/CFT inspired  relation to  massless 
  higher spin partition function in AdS$_7$ 
 \cite{Giombi:2013yva} (see  also \ci{Giombi:2014iua,Beccaria:2014xda}). 
The  latter method was generalized to   arbitrary $SO(2,6)$  representations in \cite{Beccaria:2014qea}.

Here  we will apply the  results  of \cite{Beccaria:2014qea} 
to present  the   explicit  expressions for  the a-anomaly coefficient 
(and also  the  Casimir energy $E_c$ on $S^5$)    for several types of  non-unitary  6d   conformal fields   including   some  
low-spin  supermultiplets   as well as   the  (2,0)   conformal supergravity multiplet.
 In particular,  
\begin{enumerate}
\item[i)] We will  find  that  the value of the a-anomaly coefficient  
indirectly  attributed in \cite{Cordova:2015fha}  (on the basis of  R-symmetry and gravitational  't Hooft  anomaly matching) 
to $(1,0)$  supersymmetric 6d  vector  multiplet  $\aa_{\rm (1,0)\,  vector}$ =$- {251\ov 210} \aa_{\rm (2,0)\, tensor}$
 corresponds,  in fact,  to
higher-derivative (non-unitary)   superconformal  vector multiplet  $V^{(1,0)}$  with  the Lagrangian
\footnote{The non-linear action  for  non-abelian  version of this multiplet 
was  constructed  in \cite{Ivanov:2005qf,Ivanov:2005kz}. 
It is not superconformally  invariant at  the 
quantum level since the   gauge  coupling 
$\beta$-function does not vanish (and also has chiral gauge anomaly \ci{Smilga:2006ax}). Here we consider only the free  multiplet.}
\be
\label{1.2}
\mathscr L_{V^{(1,0)}} \sim   F_{\m\n} \,\partial^{2}\, F_{\m\n}  + \overline\psi\slashed{\partial}^{3}\psi + \varphi\,
\partial^{2}\,\varphi \  . 
\ee
The   higher-derivative  model  $F_{\m\n} \,\partial^{2}\, F_{\m\n}$ 
having standard  vector gauge invariance is the $s=1$ member  of the  conformal higher spin   family  
\ci{Fradkin:1985am} in $d=6$ with the kinetic  operators
$\del^{2s+d-4}= \del^{2s+2}$. It represent   ``{massless}''  
 conformal field  
  which is  different from the  ``{massive}"  (non gauge invariant) one 
that  has 2-derivative kinetic term  which  is 
 the $s=1$ member of the family   considered in  \ci{Erdmenger:1997wy} 
 (cf. also \ci{ElShowk:2011gz}   and  below).
One could  think of \rf{1.2}  as a  UV completion of  the standard (2-derivative)   scale-invariant  but not  conformally
invariant $(1,0)$ Maxwell multiplet in 6d, though  the direct  relevance of this non-unitary 
UV theory in the context of  IR  RG flow of a-anomaly discussed  in  \cite{Cordova:2015fha} 
is unclear.  
There is also a tentative  connection to non-abelian tensor model of   \ci{Samtleben:2011fj} 
containing 3-form field: 
in 6d  the 4-derivative   vector  is dual
 to non-dynamical  3-form field (with conformally invariant 
 kinetic term $(C^\perp_3)^2$, see Appendix)
for  which the contribution to   the conformal anomaly comes from the ghost determinants and is thus of  ``non-unitary"   nature 
(as  in  Schwinger  model  or Einstein gravity  in 2d).

\item[ii)] We  will compute  the a-anomaly for  the maximally supersymmetric   $(2,0)$ 
  6d conformal supergravity  that has a schematic    Lagrangian  
\be  \la{12}
\mathscr L \sim    C_{\m\n\l\r} \del^2
C_{\m\n\l\r}  + \psi_\m \del^5 \psi_\m + ...   \ee
The non-linear action of this theory 
    can be  found as a local UV singular 
  part of the induced action of (2,0) tensor multiplet coupled to the conformal supergravity background. 
  As 
   the expression for the  $W_6$ term  in the conformal  anomaly \rf{1.1}   of  the (2,0) multiplet 
    takes a particular form  
        in  \rf{444}, 
  the  action of  the (2,0)   conformal supergravity  may be   interpreted  as a supersymmetric extension of ${\cal W}_6$.
 We shall observe that 
 when this theory is 
 coupled  to precisely  $26$ \   $(2,0)$   tensor
multiplets, the total 
conformal anomaly a-coefficient  
 vanishes. 
This  is the 6d counterpart  of the  cancellation of  4d conformal anomalies in the system of $\N=4$  conformal  supergravity 
coupled to four    $\N=4$   vector multiplets \ci{Fradkin:1983tg, Fradkin:1985am}.\foot{While we will 
 compute  the $\aa$-coefficient  just for the  free field   multiplet, 
{\em i.e.} the 
1-loop contribution,   in the  maximally supersymmetric  case 
   it is likely to be exact (as in the 4d case).}
This  cancellation is curious in view of the
following observation. Taking 5 of the 26  tensor multiplets  to be ghost-like  and
spontaneously  breaking the superconformal symmetry (and dropping higher
derivative terms, {\em i.e.} considering an IR limit)   one ends up, following  \ci{Bergshoeff:1999db},  with
a theory of the remaining $26-5=21$   tensor multiplets  coupled to
the  chiral $(2,0)$   6d  Poincar\'e supergravity. 
The latter theory  is known   to be special:
it is   gravitational  anomaly free  and  results upon  compactification
of type IIB supergravity on K3  \cite{Townsend:1983xt,Witten:1995em}.
\end{enumerate}

We shall start in section \ref{sec:2} with a brief review of the  conformal fields in 6d and present 
the  general  a-anomaly  expression derived in \ci{Beccaria:2014qea} using the  technical tools 
of  AdS$_7$/CFT$_6$
connection. 
In section \ref{sec:3}  we will consider several  unitary and non-unitary   superconformal  multiplets 
involving conformal fields of 
low spin (scalars, spinors, vectors and 2nd rank antisymmetric tensors)
and present the results for the corresponding a-anomaly and Casimir energy.
We will   study, in particular,  the non-unitary higher-derivative $(1,0)$ supermultiplet  \rf{1.2}.
In section \ref{sec:4}  we will consider the maximal  6d (2,0) 
conformal supergravity 
naturally associated with  
7d maximal  gauged supergravity with AdS$_{7}$ vacuum.
We will   demonstrate that,  when coupled to 26  (2,0) tensor multiplets, this theory has 
vanishing a-anomaly  (and  vanishing Casimir energy) 
  and discuss  some  interpretations of this fact. 
In Appendix \ref{app:s1d6}  we will 
compare  the $S^6$ partition functions for  various 6d  conformal  fields 
and also    present  the  direct   6d  derivation of the a-anomaly
for  the   non-unitary  conformal  vector field  with 2-derivative but not gauge-invariant   action
and its 4-derivative  gauge-invariant   counterpart. 

\section{6d conformal fields  and  a-anomaly from AdS$_7$}
\la{sec:2}

6d conformal fields   correspond to 
 $SO(2,6)$  conformal group  representations that will be denoted 
as  $(\Delta;\, \mathbf{h})$ where $\mathbf{h} = (h_{1},h_{2},h_{3})$  are the 
 $SO(6)$ highest weights or Young tableau  labels  ($h_i$ are all integers or  all half-integers
with $h_{1}\ge h_{2}\ge |h_{3}|$).  The dimension $\dd(\mathbf{h})$  of the $SO(6)$   representation $\mathbf{h}$ is 
\be \no 
\label{2.4}
\dd(\mathbf{h}) =\frac{1}{12}(1+h_{1}-h_{2})(1+h_{2}-h_{3})(1+h_{2}+h_{3})(2+h_{1}-h_{3})
(2+h_{1}+h_{3})(3+h_{1}+h_{2}).
\ee
The unitary irreducible representations of $SO(2,6)$ fall into four classes \cite{Metsaev:1995re,Dolan:2005wy}
\be
\label{2.2} 
\begin{split}
(i) \ \ \Delta & \ge \Delta = h_{1}+4, \ 
        \quad \text{}\  
        h_{1}>h_{2} \geq |h_3|;      \qquad 
(ii) \ \ \Delta  \ge \Delta = h_{1}+3, \  
\quad \text{}\ h_{1}  = h_2  > |h_{3}|; \\
(iii) \ \  \Delta & \ge \Delta = h_{1}+2, 
\quad \text{}\ h_{1} = h_{2} = \pm h_{3}; \qquad 
(iv) \ \ \Delta  \ge 2 \ \text{or}\  \Delta=0\ , \quad    \text{}\  h_1=h_2=h_3= 0.
\end{split}
\ee
Generic representations are non-degenerate (or ``massive''), while 
representations at the unitarity bounds are  maximally degenerate (they 
correspond, in particular, to conformal higher spins 
associated with  massless  higher spin fields in AdS$_{7}$ \cite{Metsaev:1995re}).\foot{
Their characters can be written by suitable subtractions in terms of the massive (generic) representation character 
$
\widehat{\Z}(\Delta;\, h_1,h_2,h_3) = \dd(\mathbf{h})\,\frac{q^{\Delta}}{(1-q)^{6}}\ .
$
For example, in the $\Delta = h_{1}+4$  case (i), we have the following massless character
(see \cite{Beccaria:2014qea}  for details)
$ {\Z}(h_{1}+4;\, h_1, h_2,h_3) = \widehat{\Z}(h_{1}+4;\, h_1,h_2,h_3)- \widehat{\Z}(h_{1}+5;\,h_{1}-1,h_{2},h_{3}) \ .$
}
There are also intermediate  cases of non-unitary conformal fields of non-maximal depth related to partially massless 
fields in AdS$_{7}$  (see \ci{Shaynkman:2004vu,Bekaert:2013zya,Barnich:2015tma}    for general discussions).\foot{In general, the origin of   non-unitary  of a  free conformal theory   may be 
  due to higher derivative  kinetic term  (as in conformal higher spin  field case) 
  and/or  reduced gauge invariance that does not allow  to eliminate all ghost-like components.}

Assuming   a  free conformal field  $\vp$ with
a local field theory  description with a free  action $S= \int d^d x \vp \del^n \vp$ ,
its canonical dimension  should   be  $\Delta = {d-n\ov 2}$.  
For example,   a conformal higher spin 
field  represented by a symmetric rank $s$ tensor (i.e.  $\mathbf{h}=(s,0,0)$)  with action that  has maximal gauge invariance
 has   $n= 2s + d-4$, {\em i.e.}  has  canonical $\Delta= 2-s$  and thus is unitary in $d=6$ only for $s=0$  (cf. \rf{2.2}).
 
 To  find  the corresponding conformal anomaly,  one may  couple $\vp$ to a background   metric $g_{\m\n}$ 
 (getting a classically Weyl-invariant action with $\vp$ transforming with an appropriate Weyl weight)  and 
 compute  the trace of the variation of  the    1-loop effective action 
 $\G = - \log Z[g]$   over $g_{\m\n}$.  Equivalently, to extract the a-coefficient it is sufficient to find the logarithmic UV divergent part 
 of  $\G$  computed on  6-sphere
\be 
\te \G= - B_6 \log (r \Lambda)  +... \  , \ \ \ \ \ \qquad    
B_6= {1 \ov (4 \pi)^3} \Omega(S^6)\,   b_6= {1 \ov 60} b_6 = - 96 \aa \ . \la{255} \ee
Here $\Lambda$ is a UV cutoff and $r$ is the radius of $S^6$, see \cite{Tseytlin:2013fca}  for details.

 This  direct 6d  computation of the a-coefficient    appears to require a case-by-case analysis,  but 
 there exist a remarkably  universal   method  of computing $\aa$-anomaly 
   using  the  AdS/CFT motivated    relation 
 between the determinants  of   a 
 2nd order   operator in  AdS$_{d+1}$  space  and   of the associated  
 conformal field   operator on  $d$-dimensional boundary.\foot{This relation  has  a   
 { kinematic} origin   and belongs to a general class of bulk-boundary relations discussed in 
 \ci{Barvinsky:2005ms,Barvinsky:2014kta}.
 Its AdS/CFT interpretation involves  the  bulk  counterpart  
 of  a ``{double trace}''  deformation  of the boundary CFT  (see, in particular,  
   \ci{Gubser:2002vv,Diaz:2007an,Diaz:2008hy} for scalar operators).
   In \ci{Giombi:2013yva,Tseytlin:2013fca,Giombi:2014iua}  it was  applied to the computation of the a-coefficient of  totally-symmetric 
   higher-spin conformal fields
   and  in \ci{Beccaria:2014xda,Beccaria:2014qea} it was  generalized to arbitrary conformal representations in 4d and 6d  (see
     also  \ci{Tseytlin:2013jya,Beccaria:2014jxa,Beccaria:2014xda}  for related general discussions).}
  While being a just a technical   device   (leading  to the same results 
 as the 6d computation,  as one can check on particular examples), this  method   makes full use of the underlying conformal symmetry 
 and  allows one  to compute  the  a-coefficient  for a generic  
 representation $(\Delta;\, h_1,h_2,h_3)$.

   A  conformal field $\vp$ in $\mathbb R^d$ 
    of  canonical  dimension $\D_-$  may be interpreted as a shadow (or source) field
   associated to  another  conformal field $J$  (or {\em current}) 
     of dimension $\D_+ = d-\D_-$ that  has 
     the same $SO(d)$ representation  labels  $\mathbf{h}$. For example, 
   in the context of vectorial AdS/CFT,  the current $J$  may be interpreted as a bilinear  in complex scalars which is dual to a massless
   higher spin field $\p$  in AdS$_{d+1}$  transforming 
    in the same  representation of  $SO(2,d)$ (the isometry 
group of AdS$_{d+1}$)  as $J$, i.e. $(\D_+; \mathbf{h})$.

    For a generic    field $\vp$, the associated 
    { current} needs not be conserved
   and the dual AdS$_{d+1}$  field $\p$ is  massive, {\em i.e.} its  action 
   $\sim \int d^{d+1} x \sqrt g\  \p ( - \nabla^2 + m^2) \p$ has no gauge invariance. Considering  the ratio of  determinants of the kinetic operator of $\p$  with Dirichlet (+) and Neumann (-)   boundary conditions  one can then argue 
   (cf.  \ci{Barvinsky:2005ms,Gubser:2002vv,Diaz:2007an,Diaz:2008hy,Tseytlin:2013jya})  that their ratio   should be 
   related to the determinant of the kinetic operator of the boundary conformal field $\vp$, {\em i.e.} 
   the corresponding  1-loop partition functions   should be related   as 
    \be
 \label{2.1}
Z_{{\vp}}(\mc M^d) = \frac{Z^{-}_{{\p}}(\text{AdS}_{d+1})}{Z^{+}_{{\p}}(\text{AdS}_{d+1})}\ ,
\ee
where $\mc M^{d}$  is the boundary of  $\text{AdS}_{d+1}$ ({\em e.g.},  $S^d$). 
The same relation applies also  in the {\em reducible}  cases  with gauge invariance, {\em e.g.}
the partition  function of a conformal higher spin  field $\vp_s$   is related to the ratio of  the $\pm$ partition  functions   of  the $\text{AdS}_{d+1}$ massless
higher spin field $\p_s$.  
In the even $d$  case  $Z_\p(\text{AdS}_{d+1})$  is  UV finite but  logarithmically IR divergent (due to the AdS volume factor), 
while $Z_{{\vp}}(\mc M^d) $ is IR  finite but logarithmically UV divergent as in \rf{255}. 
Identifying the two  cutoffs  allows one to find the a-coefficient for $\vp$   
by   computing  the  field $\p$ determinants in $\text{AdS}_{d+1}$  \ci{Giombi:2013yva,Tseytlin:2013fca,Giombi:2014iua,Beccaria:2014xda,Beccaria:2014qea}. 
As a result,  the a-coefficient for $\vp$ is 
$\aa= \aa^+ - \aa^-$  where $\aa^+ = f(\D_+, \mathbf{h}), \ \aa^- = f(\D_-, \mathbf{h})$.
As  it turns out, $f(\D) =- f(d-\D)$, {\em i.e.} $\aa^-=- \aa^+$, so that 
\be 
\aa(\D;\mathbf{h}) =-2  \aa^+(\D, \mathbf{h})  \ , \ \ \ \ \ \ \ \ \ \ \ \ \qquad     \D\equiv \D_+= d-\D_- \ . \la{213}\ee
In what follows we shall follow \ci{Beccaria:2014xda,Beccaria:2014qea}  and label the   conformal 
field   representation  not by its canonical dimension $\D_-$ but by the dimension 
$ \D_+= d-\D_- $ of the dual AdS  field.
The corresponding a-coefficients for fields of dimension $\D$ and $d-\D$  differ  only  by the overall sign, see \rf{2.5} below
where $\D$ will also  stand
 for $\D_+$ (the  discussion of  (non)unitarity     of a given conformal field   should  of course  be  based on its canonical dimension $\D_-$).


\iffa 
Let us briefly recall the main ideas that allows to exploit the AdS$_{d+1}$/CFT$_{d}$ framework 
in order to evaluate conformal anomalies and related quantities. 
The crucial observation is that  there are {\em kinematic}   relations   based   on the symmetries and 
 special  properties of  AdS   spaces. The prototypical example involves   the
 singlet sector of  a free   CFT$_{d}$,   its dual  higher spin   theory in AdS$_{d+1}$
 and  a {\em shadow} conformal higher spin theory in $d$  dimensions  \cite{Giombi:2013yva,Giombi:2013fka,Tseytlin:2013jya,Tseytlin:2013fca, Giombi:2014iua,Giombi:2014yra,Beccaria:2014jxa}.
 Starting, {\em e.g.},  with a   free massless   complex scalar  theory   $ \int d^d x \,   \Phi^*_r \del^2  \Phi_r $  
 one  gets   a tower of    conserved  symmetric traceless  higher spin currents  
 $J_s \sim  \Phi^*_r  \del^{s}   \Phi_r$  that are 
 primary  conformal fields of dimension $\Delta=d-2+ s \equiv \Delta_+ $. When
these  currents are added to the  to  the action with the  source shadow  fields $\vp_s(x)$  
  one notices that   $\vp_s$ has the   
 same   dimension   $\Delta_-= d-\Delta_+= 2-s$ and   the same (algebraic and gauge)
 symmetries   as   the conformal higher  spin (CHS)   fields.   
If we integrate out the  free fields $\Phi_r$ , we obtain 
an action for $\vp_s$   with  kinetic term $\KK(x,x')  \sim \langle  J_s (x) J_s  (x')   \rangle $. The  
 leading logarithmically divergent      part of  this  action  is  local and coincides with the  
 CHS  action   $\int d^d x \ \vp_s  \del^{2s+d-4} \vp_s + ... $.
 From the  dual \ads   perspective  (implying matching between  the correlators of currents  and amplitudes 
 for dual AdS fields $\p_s$)  
  this induced action  can   be  found upon the  substitution 
 of the solution of the Dirichlet problem  with $\p_s\big|_{\del} =\vp_s$ 
 into   the  classical  5d action for  a massless  spin $s$ field  $\p_s$. 
 
 This classical {\em tree-level}  relation between $d+1$ dimensional  fields $\p_s$ and  and $d$  dimensional 
  conformal higher spin   fields  $\vp_s$ 
can be extended to the corresponding one-loop   partition  functions, that is the   determinant  of the $d$ dimensional  
 kinetic operator $\KK$, 
 and  the ratio  of  determinants of     2nd-order   $d+1$ dimensional  operators  for  the field $\p_s$ 
 with Neumann-type $(\De_-$)    and Dirichlet-type ($\De_+$)   boundary conditions. 
 This relation  has  a   {\em kinematic} origin  and belongs to a general class of bulk-boundary relations 
 \ci{Barvinsky:2005ms} (see in particular \ci{Diaz:2007an,Diaz:2008hy} for scalar operators).
 Its AdS/CFT interpretation is in terms of the bulk  counterpart 
 of  a {\em double trace}  deformation  of the boundary CFT
  (see \ci{Witten:2001ua,Gubser:2002zh,Gubser:2002vv,Hartman:2006dy,Diaz:2007an,Diaz:2008hy,Giombi:2013yva}). 
  
  As discussed in details in \cite{Beccaria:2014xda}, the story can be generalized to a large extend. One can consider
  instead of a $d$ dimensional CHS field a generic primary  conformal  field 
 that will  be associated to a particular (in general, massive or massless higher spin)  field in AdS
  effectively encoding its 
 quantum characteristics. The general (1-loop) relation will thus be (for each field)
 \be
 \label{2.1}
Z_{\text{CHS}}(\mc M^{d}) = \frac{Z^{-}_{\text{HS}}(\text{AdS}_{d+1})}{Z^{+}_{\text{HS}}(\text{AdS}_{d+1})},
\ee
where $\pm$ denotes Dirichlet or Neumann boundary conditions, and $\mc M^{d} = \partial\text{AdS}_{d+1}$.
Each HS field will transform in a representation of the $SO(2,d)$ isometry 
group of AdS$_{d+1}$ ($d$ even) with labels $(\Delta; h_{1}, \dots, h_{\frac{d}{2}})$ where, in bosonic case, 
 $h_{i}$ is the length of the $i$-th row of a $SO(d)$ Young tableau. The set $(h_{1}, \dots , h_{\frac{d}{2}})$ 
 characterizes the spin (internal) representation of the field. The associated conformal field 
 will have canonical dimension $\Delta_{-} = d-\Delta$, and same spin representation.
 
 Gauge symmetries of the CHS theory involve ghost fields. They also follow the above pattern. At the group-theoretic
 level, their $SO(2,d)$ representations subtract unphysical states in the partition function and realize at the 
 field level group-theoretical quotients \cite{Beccaria:2014jxa}. Unitarity of CHS is thus the same as 
 unitarity of these representation modules. 
 
 Our main concern in this paper will be the study of generically non-unitary conformal theories in 6d by the above methods
 and non-unitarity will  be due to higher derivatives kinetic terms and/or partially reduced gauge invariance
 \cite{Bekaert:2013zya}.
\fi 

In particular, in the $d=6$  case  one  finds  for a generic massive $SO(2,6)$  representation 
\cite{Beccaria:2014qea}   (${\bf{h}}=(h_1,h_2,h_3)$, \ $\bh= h_1 + h_2 + h_3$) 
\ba\no
& \aa(\Delta;{\bf{h}}) = -\frac{(-1)^{2\bh} \dd(\bfh)}{96\times37800} \, (\Delta -3)
\Big[15 (\Delta -3)^6 \\
&  \qquad \qquad \qquad   -21 (\Delta -3)^4 \left[h_3^2+h_1 \left(h_1+4\right)+h_2 \left(h_2+2\right)+5\right]\no 
\\ \no 
&   \qquad \qquad \qquad 
 +  35 (\Delta -3)^2 \big[\left(h_1+2\right)^2
   \left(h_2+1\right)^2+\big(h_1 \left(h_1+4\right)+h_2 \left(h_2+2\right)+5\big) h_3^2\big] \\
   &    \qquad \qquad \qquad   
   - 105  \left(h_1+2\right)^2
   \left(h_2+1\right)^2 h_3^2\Big] \ .  \label{2.5}
\end{align}
In the case of  degenerate  representations  ({\em e.g.}, short ones saturating a unitarity bound),  one needs to  
 combine the corresponding massive representation expressions  appropriately 
 ({\em i.e.}   subtract ``{ghost}''  contributions).  

   One   may also  apply a similar method to    compute   the 
  Casimir energy on $S^{d-1}$   by using \rf{2.1}  in the case of
  the boundary   being  $\mathbb R \times S^{d-1}$ \ci{Giombi:2014yra,
  Beccaria:2014jxa,Beccaria:2014xda}. 
  The  general expression for the Casimir energy on $S^5$  
  of  a 6d   conformal field  is found to be 
  \cite{Beccaria:2014qea} 
   \be
 \label{2.6}
E_{c}(\Delta;\, \mathbf{h}) = - \frac{(-1)^{2\bh}\dd(\mathbf{h})\,}{60480}\,
(\Delta-3)\, \Big[
12\,(\Delta-3)^{6}-126\,(\Delta-3)^{4}+336\,(\Delta-3)^{2}-191
\Big].
\ee
Note that  the expression for $E_c$ is (in contrast to the one for $\aa$ in \rf{2.5})  
 scheme  dependent   in general: it is  determined  by  the  $\aa$-coefficient  
but also by  the scheme-dependent  coefficients of the total derivative terms
in  $D_6$ in \rf{1.1} 
\cite{Cappelli:1988vw}. Here we use a particular (heat kernel or $\zeta$-function) scheme 
in which $\aa$ and $E_c$ are not simply proportional  (cf.  \cite{Herzog:2013ed})  so  that 
 the computation of $E_c$ provides an independent information  about 
the $D_6$ terms in \rf{1.1}.


\section{Unitary and non-unitary { low spin}  6d  conformal  multiplets}
\la{sec:3}

Let us consider generic  (higher derivative)  6d conformal theories  for a  scalar ($\varphi$), 
spin $\frac{1}{2}$ fermion ($\psi$), vector ($V$), and 2nd rank antisymmetric tensor  ($T$).  
We adopt a notation that displays the order of derivatives in  the corresponding kinetic term 
\be
\la{3.1}
\varphi^{(n)}\rightarrow \varphi\,\partial^{n}\,\varphi, \qquad
\psi^{(n)}\rightarrow \psi\,\slashed{\partial}^{n}\,\psi, \qquad 
V^{(n)}\rightarrow V_\m \,\partial^{n}\,V_{\m}, \qquad T^{(n)}\rightarrow T_{\m\n} \,\partial^{n}\,T_{\m\n} \ .   
\ee
In this section  we shall focus on  particular cases 
$\varphi\equiv \varphi ^{(2)}$, $\varphi ^{(4)}$; \ $\psi\equiv \psi ^{(1)}$, $\psi ^{(3)}$;\ \
 $V^{(2)}$, $V^{(4)}$,  and the standard $T\equiv T^{(2)} $ with 
 self-dual field strength $H$.\foot{As usual,  when talking about actions for self-dual antisymmetric  tensor fields 
 we will  be assuming that  self-duality condition is relaxed and imposed on equations of motion or in relevant quantum  computation.} 

In $d=6$  the conformal  field $V^{(2)}$  has canonical dimension $\D_- = 2$ 
 ({\em i.e.} the corresponding dual representation is $(4; 1,0,0)$)  
and  thus is below the 
unitarity bound in \rf{2.2}. Indeed, it is not described  by  the usual  Maxwell Lagrangian $\sim F_{\m\n} F_{\m\n}$
but rather  by a  non gauge-invariant one $ (\del_\m V^{(2)}_\n)^2 - { 3 \ov 2} (\del_\m V^{(2)}_\m)^2$ (see Appendix \ref{app:s1d6})  which is a special $s=1$ case of 
a class of 2nd derivative conformal  spin $s$ field  actions 
discussed in \cite{Erdmenger:1997wy}. Indeed, the Maxwell vector  is scale invariant  but not 
conformally invariant in $d=6$ (see, {\em e.g.},  \cite{ElShowk:2011gz}). 

At the same  time, 
 the field $V^{(4)}$  is described  by  the gauge
  invariant conformal theory  with  the  kinetic term $\sim F_{\m\n}\partial^{2}F_{\m\n}$    where $F_{\m\n} = \del_\m V^{(4)}_\n 
  - \del_\n V^{(4)}_\m$. This is 
    the $s=1$  member  of the  conformal higher spin   family in $d=6$ 
    \cite{Fradkin:1985am,Tseytlin:2013jya,Tseytlin:2013fca}.  
    Its  canonical dimension $\D_- = 1$, {\em i.e.} it is also non-unitary 
    (cf.   \rf{2.2})
  but now  the non-unitarity  may be attributed   to   its   higher-derivative kinetic term. 
  The corresponding Weyl-invariant action on curved background is presented in Appendix (see  \rf{A.101}).

The count of  dynamical ({on-shell}) degrees of freedom $\nu$ for these  fields goes as follows.
Since for the 2-derivative scalar  $\nu(\vp) =1$ we   have  $\nu(\varphi^{(n)})=n/2$. 
For 6d Majorana-Weyl fermion  $\psi^{(n)}$  
we get  $\nu(\psi^{(n)})=-2n$. 
The  standard gauge-invariant antisymmetric tensor $T_{\m\n}$   with selfdual  field 
strength  has $\nu(T) = 3$. 
The conformal vector $V^{(n)}$
with $n\neq 4$ has no gauge invariance  
so  $\nu(V^{(n\neq 4)}) = d\times n/2 = 3n$. The case $n=4$ in 6d 
is special:  because  of gauge invariance  the action is $V_{\m}^{\perp}\,\Box^{2}\,V_\m^{\perp}$
and there is  also a  factor $(\det\Box)^{1/2}$   coming from the measure (after one sets $V_\m = V_\m^{\perp} + \del_\m \sigma$ 
and divides over  the gauge group  volume). As a result, 
$\nu(V^{(4)}) = (6-1)\times 2 -1 = 9$. 

Applying \rf{2.5},\rf{2.6} one can compute the  a-anomaly  and the Casimir energy   corresponding  to these conformal fields. 
Table~\ref{T1} lists the (dual or AdS-adapted) representation content,  
on-shell degrees of freeedom, a-anomaly, and  
Casimir energy of these 6d conformal fields.
Note  that  the value for the a-anomaly
  for $V^{(4)}$ is indeed  the same  as for $s=1$ 
 conformal higher spin field  in 6d  found in \ci{Tseytlin:2013fca}, see   also Appendix.
\begin{table}[H]
\be
\def\arraystretch{1.3}
\begin{array}{|c|c|c|c|c|}
\hline
\mbox{field} & SO(2,6) & \nu &  7! \,\aa & 7!\, E_{c} \\
\hline
\varphi
 & (4; 0,0,0) &  1 & -\frac{5}{72} & -\frac{31}{12} \\
\hline
\varphi^{(4)} & (5;0,0,0) & 2 & \frac{4}{9} & \frac{95}{6} \\
\hline
\psi & (\frac{7}{2}; \frac{1}{2}, \frac{1}{2}, \frac{1}{2}) &  -2 & -\frac{191}{288} & -\frac{1835}{96}\\
\hline
\psi^{(3)} & (\frac{9}{2}; \frac{1}{2}, \frac{1}{2}, \frac{1}{2}) & -6 & \frac{39}{32} & \frac{1021}{32}\\
\hline
V^{(2)} & (4; 1,0,0) & 6 & -1 & -\frac{31}{2}\\
\hline
V^{(4)} & (5; 1,0,0)-(6; 0,0,0) & 9 & \frac{275}{8} & \frac{1755}{4} \\
\hline
T & \frac{1}{2}[(4;1,1,0)-(5;1,0,0)+(6;0,0,0)] &3 & -\frac{221}{8} & -\frac{955}{4}  \\
\hline
\end{array}\notag
\ee
\caption{Some  6d  conformal fields   and their  properties. 
Here $SO(2,6)$ quantum numbers $(\Delta; h_1, h_2, h_3)$  refer to the 
dual   field, {\em i.e.} 
  $\Delta= 6 - \Delta_-$ where $\Delta_-$ is canonical dimension 
of a conformal  field.  Listed  are the   numbers of  dynamical d.o.f. $\nu$   and the values of 
 the  a-anomaly  and the Casimir energy $E_c$ on $S^5$. Combinations of representations account for shortening (gauge freedom). 
 Here $T$  has  self-dual field strength  (which is  indicated   by  1/2 in representation content).  }
\label{T1}
\end{table}
These fields  may   be combined into supermultiplets 
with zero total   number of degrees of freedom $\nu=0$. 
In particular, we  may  consider 
$(1,0)$  hyper, tensor and vector superconformal multiplets  
\be
\label{3.2}
S^{(1,0)} = 4\varphi+2\psi, \qquad \quad 
T^{(1,0)} =  \varphi+2\psi+ T, \qquad 
V^{(1,0)} = 3\varphi+2\psi^{(3)}+V^{(4)} \ , 
\ee
as well as the  (2,0)   tensor multiplet 
\be
\la{3.4}
T^{(2,0)} = T^{(1,0)}  + S^{(1,0)}= 5\varphi+4\psi+T \ .
\ee
While the unitary $S^{(1,0)}$,  $T^{(1,0)}$  and $T^{(2,0)}$  are familiar, 
the non-unitary multiplet $V^{(1,0)}$ may  be less so. 
Its  Lagrangian \rf{1.2}   is essentially  like the standard Maxwell supermultiplet Lagrangian  with 
an extra $\del^2$   operator in the kinetic terms (see \ci{Ivanov:2005qf}).

Expressing  the a-anomaly and the Casimir energy  for these multiplets in terms of their 
values   for the  (2,0)   tensor multiplet found in \cite{Bastianelli:2000hi,Beccaria:2014qea}
we obtain 
\be
\la{3.5}
\begin{split}
& \aa(S^{(1,0)}, T^{(1,0)}, V^{(1,0)}) = 
\left(\frac{11}{210}, \frac{199}{210}, -\frac{251}{210}\right)\,\aa(T^{(2,0)}),\\
& E_{c}(S^{(1,0)}, T^{(1,0)}, V^{(1,0)}) = 
\left(\frac{37}{250}, \frac{213}{250}, -\frac{377}{250}\right)\,E_{c}(T^{(2,0)}), \\
&
 \aa(T^{(2,0)}) = -\frac{7}{1152}, \qquad\qquad  E_{c}(T^{(2,0)}) = -\frac{25}{384}\ .
\end{split}
\ee
We observe   that the  value $-\frac{251}{210} \,\aa(T^{(2,0)})$ attributed (on the basis of 't Hooft anomaly matching) in \ci{Cordova:2015fha} 
 to 
the   standard (non-conformal)    (1,0)  Maxwell multiplet   corresponds, in fact, 
 to  the non-unitary 
higher-derivative $V^{(1,0)}$  multiplet.

Let us  make few comments  to try to understand  this coincidence
(see also a discussion of non-unitary $V^{(4)}$ field and its 3-form dual
  in Appendix).  
An important  point   should  be that  non-conformal  Maxwell vector multiplet  should  be emerging upon spontaneous 
breaking  of conformal invariance  from a conformal system of interacting  vector and tensor  multiplets 
with $ \vp F_{\m\n} F^{\m\n}$  term in the Lagrangian \ci{Samtleben:2011fj} 
where $\vp$  is a scalar of tensor multiplet 
  that has dimension 2   in 6d.  
    That means this  vector  has canonical dimension 1 as  for 4-derivative vector  field 
   and that changes also   the assignment of R-charge to the corresponding fermion  and then the 
   count of anomalies and thus of a-anomaly should go as in $V^{(1,0)}$ multiplet case.  
Starting with non-conformal Maxwell (1,0) 6d multiplet  with symbolic Lagrangian 
 $m^2 ( F_{\m\n} F_{\m\n}  +  \psi \del \psi +   D^2)$  (where $m^2 = < \vp >$) one may embed   it into a  higher-derivative theory 
 $ F_{\m\n} ( m^2 + \del^2)  F_{\m\n}  + \psi   ( m^2 + \del^2) \del \psi +   D ( m^2 + \del^2)  D$
 which is  conformal in the UV. The  axial  anomalies will be the same (as $\del$ and $\del^3$ fermions
 have the same chiral anomalies, see, e.g., \ci{Smilga:2006ax}) but  a-anomaly  will be controlled 
 by the conformal UV theory, i.e. will be the same as   for the $V^{(1,0)}$ multiplet in \rf{3.2}. 
 
 A similar  example  (suggested in \ci{Cordova:2015fha})  exists in 4 dimensions:  
 starting 
 with  a  non-conformal linear   multiplet $(B_{\m\n},\vp,  \psi, f)$  one can  dualize  the antisymmetric tensor\foot{4d theory of 
 gauge-invariant antisymmetric   tensor $H_{\m\n\l} H^{\m\n\l}$   is not 
 conformal already at the classical level, so the notion of   conformal anomaly   does not apply. 
 One may still formally consider the corresponding  $B_4$  coefficient of logarithmic UV divergence of partition function 
 on  a curved background
 (e.g., on $S^4$)  and thus  define the ``analog"  of the one-loop a-coefficient (cf. \rf{255}); its  value will be preserved by the 
 duality 
 transformation  in the path integral (see, e.g., \ci{Grisaru:1984vk,Fradkin:1984ai}).}
  into a scalar 
  getting a non-conformal  analog of the 
   chiral multiplet   
   (with the dual scalar  shift symmetry  prohibiting $R \vp^2$   term  on a curved  background)  where 
 the  $U(1)_R$ charge  of the scalar should be 0 and that of the associated fermion being -1. 
  If one then formally applies the relation between the  R-charge and a-anomaly coefficient which is valid in superconformal 
  $\N=1$ theories   one   finds that the associated   conformal anomaly coefficient should be 
  $\aa_{\rm  d=4}= - { 3 \ov 16}$    \ci{Cordova:2015fha}.  In view  of the above discussion we should  expect that this value should  actually correspond 
  to the  conformally invariant  higher-derivative version \ci{Ferrara:1977mv}  of the chiral  multiplet with  extra $\del^2$ in  kinetic terms, i.e. 
  with  the action $\int d^4 x d^4 \theta \bar \Phi \del^2 \Phi \to \int d^4 x (
  \vp^* \del^4 \vp + \psi \del^3 \psi  + f^* \del^2 f)$. 
  Adding extra derivatives effectively shifts the scaling dimension and R-charge assignments  compared to the standard 
  chiral multiplet ones. 
  Indeed,  starting  with its  Weyl-invariant generalization  to curved space  and computing 
  the corresponding a-anomaly contributions    \ci{Fradkin:1981jc,Fradkin:1985am} 
  (see, e.g., Table 2 in \ci{Beccaria:2014xda}) one finds\foot{The same   result  is found by using the
  AdS$_5$  motivated count of $\aa$-anomaly   based on the analog \ci{Beccaria:2014xda}  of eq.\rf{2.5}, {\em i.e.}
  summing up the contributions of the corresponding  $SO(2,4) $ representations:

  \noindent
  $
  \aa_{\rm  d=4}=2 \aa(4;0,0) +  [ \aa({7 \ov 2} ;{1\ov 2 },0)+  \aa({7 \ov 2} ;0,{1\ov 2 })  ] + 2 \aa(3;0,0) $.}
  \be\la{356}
   \aa_{\rm  d=4} =
  2 \times ( - {7 \ov 90} )- { 3 \ov 80}   + 2 \times { 1 \ov 360} = - { 3 \ov 16} \ .\ee
   \iffa 
The point is  that   H^2  theory is scale but not conf invariant or Weyl inv on
 curved backgrnd; 
dualization can be done on curved  backgrnd but then it gives us 
dX* dX     *without*  R/6 X*X   term,   so non-Weyl-inv theory. 
Then  part function on sphere  is still ok    with some "a"  log L 
 but this "a" will not be same as a  of proper conf operator  with R/6  coupling. 
 [I suggest we look still at values of this "a" as it comes from log Z on S^6 just in case]
But theory we start with is not conformal, so here idea was 
that  right theory   is 
 dX* (m^2 +  d^2)   dX + psi  (m^2    + d^2)  d psi + F* (m^2+ d^2 ) F 
which is  non-conf in IR but conf in UV 
--  that requires assuming that  X has zero scale dimension (just a choice 
to reflect zero U1  R-charge ) 
so it seems   there is rather trivial explanation  of what they found -- 
if they assign R-charge to  V_m as they did in d=6  for Maxwell that is same 
looking at F_mn d^2 F_mn  theory...
Still, may be we can make  4d analogy more   direct -- 
what if we dualize  F_mn F_mn  in 6d: we get 
H_mnkl H_mnkl    ,    H_mnkl = d_m C_nkl + ... 
This is  again non-conf theory; but starting from it presumably assigns to F_mn 
  same   R-charge 
as in F_mn d^2 F_mn theory ...
So may be indeed   best to understand 4d example in all  detail
-- look at all a-values ... 
but still main issue  is dynamical mechanism  of how F_mn F_mn   emerges 
in 6d which is not conf inv from some conf inv system. 
B_4 of H_mnk H_mnk and of  conf coupled  scalar 
d_m X d_m X*  are not same  but I do not recall if B_4 
of them is not or is same  if we do not add conf coupling 
-- so what if we  just look at  B_4 on sphere. 
antisymm tensors were discussed e.g.   in  sect 4 of 
http://www.sciencedirect.com/science/article/pii/0550321383900226
so may be by some miracle  the non-conformal  linear multiplet itself has its B_4 coeff on S^4 matching  B_4  of conformal d^4  chiral multiplet ...
probably very unlikely... (as   Z on S4 does not play any role
in relation between a  and U1_r charge) 
but if that were to happen we could use similar logic in 6d, i.e. 
look at dual to vector multiplet   (i.e. involving C_nkl field)
\fi 
\iffa 
It is interesting to note    that  representation content of   higher derivative  vector
  $V^{(4)}$ in   Table \ref{T1} is the same  as the ``ghost`` part of the  standard  
  gauge-invariant antisymmetric tensor $T$ (relaxing  its 
  selfduality  condition for simplicity).  Given an antisymmetric tensor 
   path integral $\int [d T_{\m\n}]   \exp ( - S)$
    with the action  $S = \k \int d^6x \sqrt g\, H_{\m\n\l}^2, \   H_{\m\n\l}= \del_{[\m} T_{\n\l}]$
    for non-zero  coupling   constant 
    $\k$   the corresponding a-anomaly   will be (twice)  as of   the 
    $T$ field in Table \ref{T1},  while  for $\k=0$,  {\em  i.e.}
      with no kinetic  term and all contribution coming from gauge determinants in the measure
   we will  get minus the $V^{(4)}$ in   Table \ref{T1}. 
   Since non-abelian    (1,0)  superconformal models   
  involve  interacting  systems  of antisymmetric tensors  and vectors  
    \ci{Samtleben:2011fj}
   with scalar couplings  in vector kinetic terms (implying that the latter do not directly contribute to a-anomaly) 
    the above   observation  may  contain a clue towards 
 an   explanation of the above coincidence (cf. also discussion in Appendix). 
\fi 
\iffa 
\subsection{Combinations with vanishing $\aa$ (and $E_{c}$)}

One can look for combinations with zero degrees of freedom $\nu=0$ and vanishing anomaly 
$\aa=0$ involving the above  fields.  Focusing on combinations where each of the fields 
$\varphi, \varphi^{(4)}, \psi, \psi^{(3)}$ has multiplicity $\le 10$ and there are up to 4 tensor fields $H$,
we find the following list of six independent ({\em i.e.} we drop trivial linear combinations of simple cases)
solutions
\be
\la{3.6}
\begin{split}
X_{1} &= \varphi^{(4)}+\psi+\psi^{(3)}+V^{(2)}, \\
X_{2} &= 8\varphi+\psi+\psi^{(3)}, \\
X_{3} &= 9\psi+\psi^{(3)}+2V^{(2)}+V^{(4)}+H, \\
X_{4} &= \varphi+10\varphi^{(4)}+8\psi^{(3)}+2V^{(4)}+3H, \\
X_{5} &= 7\varphi+2\psi+10\psi^{(3)}+3V^{(2)}+3V^{(4)}+4H, \\
X_{6} &= \varphi+8\varphi^{(4)}+7\psi+7\psi^{(3)}+3V^{(4)}+4H.
\end{split}
\ee
None of these can be written as
combinations of $S^{(1,0)}$, $V^{(1,0)}$, $T^{(1,0)}$ supermultiplets and  they   have
$E_c\neq 0$.  If we also impose $E_{c}=0$, the minimal solutions appear to be more complicated. Requiring 
that each of the fields 
$\varphi, \varphi^{(4)}, \psi, \psi^{(3)}$ has multiplicity $\le 50$ and that there are again up to 4 tensor fields $H$,
we find the following list of four independent (contrived)
solutions
\be
\la{3.8}
\begin{split}
Y_{1} &= 40\varphi+3\varphi^{(4)}+8\psi^{(3)}+3V^{(2)}, \\
Y_{2} &= 34\varphi+9\varphi^{(4)}+20\psi+4\psi^{(3)}+V^{(4)}+H, \\
Y_{3} &= 42\varphi+3\varphi^{(4)}+42\psi+2\psi^{(3)}+4V^{(4)}+4H, \\
Y_{4} &= 50\varphi+8\varphi^{(4)}+30\psi+6\psi^{(3)}+V^{(2)}+2V^{(4)}+2H.
\end{split}
\ee
%
\fi 
Going back to the 6d values of a-anomaly in \rf{3.5}, for a combination 
$n_{s}\,S^{(1,0)}+n_{t}\,T^{(1,0)}+n_{v}\,V^{(1,0)}  $
of several multiplets 
we get 
\be
\la{3.10}
\begin{split}
\aa(n_{s},n_{t},n_{v}) &=\frac{1}{210} (11 n_s+199 n_t-251 n_v)\,\,\aa(T^{(2,0)}), \\
E_{c}(n_{s},n_{t},n_{v}) &=\frac{1}{250} \left(37 n_s+213 n_t-377 n_v\right)\,\,E_{c}(T^{(2,0)}).
\end{split}
\ee
Since the non-unitary  vector multiplet  gives  a negative   contribution as compared to the two  unitary  ones
it is of interest to  see if the a-anomaly   cancellation   condition 
 $\aa(n_{s},n_{t},n_{v})=0$  has  simple solutions for  integers $n_{s},n_{t},n_{v}$. 
 We find  (here $q_{i}$ are non-negative integers) 
\be\no 
\la{3.11}
\begin{split}
n_{s} &= 251 q_1+96 q_2+37   q_3+15 q_4+8 q_5+q_6, \qquad 
n_{t} = q_2+3 q_3+8 q_4+21   q_5+34 q_6+251 q_7,\\
n_{v} &= 11 q_1+5 q_2+4 q_3+7   q_4+17 q_5+27 q_6+199  q_7.
\end{split}
\ee
Two  special cases  are
$ (n_{s},n_{t},n_{v}) = (15,8,7)$ \text{and } (8,21,17).
\foot{ If we require in addition $E_{c}(n_{s},n_{t},n_{v})=0$,
then the most general solution is 
$n_{s} = 1078 q_1, \ \ 
n_{t} = 257 q_1,\ \ 
n_{v} = 251 q_1,
$ 
where $q_{1}$ is a non-negative integer.}

\section{(2,0) Conformal Supergravity in 6d }
\label{sec:4}

Let us now  consider an example  of  higher spin  6d supermultiplet --   maximally extended  (2,0) 
conformal supergravity (CSG). While as a background  (off-shell)   multiplet  coupled to   (2,0) 
tensor multiplet  it was  constructed earlier   in \cite{Bergshoeff:1999db}, 
the corresponding action for dynamical CSG fields  was  not discussed  in the past. 

The strategy to  determine  this action  may  be the same as in 4d case   where  the 
action of $\N=4$  CSG  can be found as an {\em induced} one: 
  either as a local   UV singular part  of the  one-loop 
  effective action of $\N=4$  Maxwell  multiplet   
 coupled to $\N=4$ CSG  background 
   or as an IR  singular part of  the  value of the $\N=8, d=5$   gauged supergravity action evaluated 
 on  solution of the corresponding Dirichlet problem \ci{Liu:1998bu,Buchbinder:2012uh}. 
 Similarly, in the $d=6$  case we may  consider either the UV   divergent 
 term   in the induced  action for $(2,0)$  tensor multiplet coupled to (2,0)  CSG   multiplet or 
 the  IR singular part  \ci{Henningson:1998gx} 
 of the value  of the  action of  maximal 7d gauged supergravity \ci{Gunaydin:1984wc,vanNieuwenhuizen:1984iz}.
 In particular, the structure of the  conformal anomaly for (2,0) tensor multiplet \ci{Bastianelli:2000hi}
   implies that 
 the  CSG  action \rf{12}   should be the supersymmetric extension of the  corresponding special 
 ${\cal W}_6$ term  in \rf{444}.\foot{This  
 special  Weyl invariant  was first  suggested as  a  ``simplest one" in   \ci{Bonora:1985cq}. It 
 can be written  only in terms of Ricci tensor  and  that is 
  why  it  came out also   in the  holographic  computation  \ci{Henningson:1998gx}
  of the  c-part  (i.e. $W_6$ in \rf{1.1}) of the conformal anomaly of strongly coupled (2,0) theory 
  (i.e. as the  logarithmic IR divergence  of the 7d Einstein  action evaluated 
  on solution with prescribed boundary metric).  Being protected by maximal 6d supersymmetry this 
  term appears  \ci{Bastianelli:2000hi} also  in the conformal anomaly of free (2,0) tensor multiplet \rf{444}.
  The fact that  the Weyl invariant ${\cal W}_6$  
  has a particular   structure 
  (it  can expressed in terms of Ricci tensor  and is at most linear in Weyl tensor if one drops a total derivative term $\sim {\cal E}_6$)  implies  that it 
  can be rewriten  as a 2nd derivative action  involving  several tensors  of rank $\leq 2$
  and it  is uniquely selected by this requirement \ci{Metsaev:2010kp}. It is also  a natural  choice for 
  the Weyl gravity action in 6 dimensions  which   shares the 4d  property   that its classical  ``S-matrix" for the physical graviton mode   in dS$_6$  (or  euclidean AdS$_6$) is the same as in the Einstein theory 
  \ci{Maldacena:2011mk}  (as this  combination ${\cal W}_6$ 
  appears also  in the  corresponding regularized  6d volume 
  in \ci{Chang:2005ska}).}
   This 
definition of the CSG action   as the local part of   the induced  theory 
guarantees the right symmetries and thus  allows in principle  to find   its full non-linear form.


\begin{table}[H]
\be
\def\arraystretch{1.3}
\begin{array}{|cc|c|cc|cc|}
\hline
\text{field} &    & SO(2,6) & USp(4) & \text{dim} &  7!\,\aa & 7!\,E_{c} \\
\hline
\text{scalar} &  D^{ij,k\ell} & (4;0,0,0) & [0,2] & 14 & -\frac{5}{72} & -\frac{31}{12} \\
\text{fermion} & \chi^{ij}_{k} &  (\frac{9}{2};\frac{1}{2},\frac{1}{2},-\frac{1}{2}) & [1,1] & 16 & 
\frac{39}{32} & \frac{1021}{32}  \\
\text{3-form} &  T^{ij}_{\m\n\l} & (5; 1,1,-1) & [0,1] & 5 & 
\frac{166}{9} & \frac{475}{3}\\
\hline 
\text{vector} & V_{\mu}^{ij} &  (5; 1,0,0)-(6;0,0,0) & [2,0]  & 10 & 
\frac{275}{8} & \frac{1755}{4}\\
\text{gravitino} & \psi_{\mu}^{i} &  (\frac{11}{2};\frac{3}{2},\frac{1}{2},-\frac{1}{2}) 
-(\frac{13}{2};\frac{1}{2},\frac{1}{2},-\frac{1}{2}) & [1,0]  & 4 & 
-\frac{4643}{16} & -\frac{137637}{16}\\
\text{graviton} &  h_{\mu\nu} & (6;2,0,0)-(7;1,0,0) & [0,0]  & 1 & 
\frac{3005}{2} & 37287\\
\hline
\end{array}
\nonumber
\ee
\caption{Representations, a-anomaly and Casimir energy for  fields of (2,0)  6d conformal supergravity. 
As in  Table \ref{T1}  the  $SO(2,6)$  representations $(\Delta; h_1,h_2,h_3)$ refer   to  dual fields 
of AdS$_7$  supergravity, {\em i.e.} $\D=6-\D_-$  where $\D_-$ is canonical dimension of the  6d
conformal field.
\label{T2}
}
\end{table} 

 To compute the associated  a-anomaly coefficient  and Casimir energy  we may use again the general expressions \rf{2.5} and \rf{2.6}.
 This   was essentially done already in \ci{Beccaria:2014qea}. 
 The  relevant field representation content is  readily  determined, using, {\em e.g.},  the   relation to the 
  maximal   gauged 7d supergravity  (with   AdS$_7$  vacuum) which is   the bottom level of the Kaluza-Klein tower of multiplets corresponding to 
   11d    supergravity compactified on $S^4$  
  (for translation  between fields of 
   7d  gauged and 6d conformal supergravities see  \ci{Bergshoeff:1999db,Nishimura:1999av}). 
\iffa 
The fields of maximal gauged  7d supergravity with AdS$_{7}$  vacuum are 
the bottom level of the Kaluza-Klein tower of 11d    supergravity compactified on $S^4$  
\ci{Casher:1984ym,Gunaydin:1984wc,vanNieuwenhuizen:1984iz}. 
According to   Sec.~\ref{sec:2}, they are naturally associated with the fields of 
$(2,0)$ 6d conformal supergravity.
\fi 
The resulting  data   is presented 
 in Table ~\ref{T2}. 
\footnote{Here $i,j,...=1,...,4$   are $USp(4)$   indices and $\m,\n,...$ are 6d  indices.
Let us note that the same values of the Casimir energy
 for these fields were  previously  found  in \cite{Gibbons:2006ij} (see also \cite{Beccaria:2014qea}).}
 
 The  number of derivatives in kinetic terms is determined by the 
dimensions of the  fields. 
 $D^{ij, k\ell}$ 
is a  conformal scalar with canonical dimension $\Delta_{-} = 6-4=2$, {\em i.e.} is  the  standard  scalar $\varphi$
discussed in Sec.~\ref{sec:3}.
$\chi_{k}^{ij}$ 
is a conformal spin $\frac{1}{2}$ fermion spinor 
with $\Delta_{-} =\frac{3}{2}$, {\em i.e.}  it is the 
3-derivative fermion $\psi^{(3)}$ in \rf{3.1}. 
 $T^{ij}_{\m\n\l}$ 
 is a non gauge invariant (anti) selfdual 
antisymmetric 3rd rank tensor 
with $\Delta_{-}= 1$, {\em i.e.}  it  should  have    kinetic term  $T \partial^{4}T$. 

The  remaining  three fields in Table~\ref{T2}  have maximal  gauge invariance, {\em i.e.}  they  are members of conformal higher spin family. The 
conformal vector  
$V^{ij}_{\mu} $ 
 with $\Delta_{-} = 1$  is of 
 $V^{(4)}$ type in \rf{3.1}.  
The  4 conformal gravitini\foot{We consider  all  6d spinors  as Majorana-Weyl, {\em i.e.} 
 there are 2 Weyl gravitini corresponding to $(2,0)$  supersymmetry.} 
  $\psi_{\mu}^{i}$  
have $\Delta_{-}=\frac{1}{2}$ and  thus the Lagrangian 
$\sim \psi_{\mu}\slashed{\partial}^5\psi_{\mu}$. 
The   graviton 
has  $\Delta_{-} = 0$, {\em i.e.} its kinetic term  has 6 derivatives
as appropriate for  the conformal gravity in 6d   ($L\sim C\, \del^2 \, C+...$, 
see, {\em e.g.},  \cite{Metsaev:2010kp}). 

 To summarize, the  symbolic  form of  the linearized  Lagrangian  
  of the  maximal (2,0)  6d conformal supergravity is  
  (here $\psi_{\m\n}$ and $F_{\m\n}$  
   are the gravitino  and the vector field strengths)\foot{The check that the total number of dynamical d.o.f. 
   vanishes goes as follows. 
 $T$ obeys  self-dual condition (the sign of its $SO(6)$   label  $h_{3}$ is fixed) giving factor  $\frac{1}{2}$; it 
has  kinetic term $\sim \Box^{2}$  and  $\frac{6\cdot 5\cdot 4}{3!}$ components  with  no
gauge invariance, therefore $\nu(T) = 20$.
For the 6d conformal graviton and gravitino,  we can  generalize the counting in \cite{Tseytlin:2013jya}.
For a bosonic conformal higher spin  field with spin $s$ in $d$ dimensions we have
$
\nu_{s} = \big(s+\frac{d-4}{2}\big)\,N_{s}-\big(s+1+\frac{d-4}{2}\big)\, N_{s-1},
$
with $N_{s} = \binom{s+d-1}{s}-\binom{s+d-3}{s-2}$.
For half-integer $s = \s+\frac{1}{2}$, we may use the same expression, but with 
$N_{\s} = q\,\big[\binom{\s+d-1}{\s}-\binom{\s+d-2}{\s-1}\big]$,
where $q= \frac{1}{2}\,2^{d/2}$    for Majorana-Weyl spinors. This gives 
$\nu(h_{\mu\nu}) = 36$ and $\nu(\psi_{\mu}) = -36$.
    Thus finally\\
\noindent 
$\nu(\text{CSG}^{(2,0)})  = 14\times 1+16\times(-6)+5\times 20+10\times 9+4\times (-36)+36=0.
$ 
}
\be
\label{4.1}
L_{\text{(2,0)  CSG } } \sim  C_{\m\n\l\r} \del^2 C_{\m\n\l\r}   +
\psi^i _{\m\n} \del^3  \psi^i_{ \m\n} +  F_{\m\n}^{ij}  \del^2 F_{\m\n}^{ij}
+ T^{ij}_{\m\n\l} \del^4 T_{\m\n\l}^{ij} 
+ \chi^{ij}_k \del^3 \chi^{ij}_k  +  D^{ij,kl} \del^2  D^{ij,kl}   \ . 
\ee
 This resembles the Lagrangian of $\N=4$  CSG in 4d    \cite{Bergshoeff:1980is,Fradkin:1985am}
 but   with extra $\del^2$   factors in the    kinetic terms (suggesting a relation via some sort of dimensional  reduction).  
 In   full  non-linear   theory the vector $V^{ij}_\m$ will be  a  $USp(4)=SO(5)$  non-abelian gauge field   
 gauging the R-symmetry   of the superconformal   group (i.e. coupled to the R-symmetry current of the (2,0) 
 tensor multiplet). 
 
The total  values of the a-anomaly and $E_c$   for the (2,0) conformal supergravity multiplet 
are  obtained by summing the contributions  of  all the fields  in Table \ref{T2}  taking into  account  their $USp(4)$ multiplicities (cf.  \rf{3.5}) 
\be
\la{4.4}
\begin{split}
\aa(\text{CSG}^{(2,0)}) = & \frac{91}{576} = -26\,\aa(T^{(2,0)})\ , \\
E_{c}(\text{CSG}^{(2,0)}) =& \frac{325}{192} = -26\,E_{c}(T^{(2,0)})\ .
\end{split}
\ee
Remarkably, these values   mean  that  (2,0)  conformal   supergravity     coupled to 26   (2,0) tensor multiplets 
has   vanishing a-anomaly and  the Casimir energy.
The fact that   the values  of a-anomaly and $E_c$ are  correlated  implies  the cancellation 
of the  total derivative $D_6$ term in \rf{1.1} in this maximally supersymmetric case.\footnote{Let us  mention   that 
   one may also consider the case of  less supersymmetric 
   $(1,0)$  6d CSG  \cite{Coomans:2011ih} whose (non-auxiliary) field content is the same 
   as in  Table \ref{T2}, but with multiplicities $D\, (1)$, $\chi(2)$, $T(1)$, $V_{\mu}(3) $, $\psi_{\mu}(2)$, $h_{\mu\nu}(1)$, 
   so that the total number of  dynamical d.o.f.   is again zero: 
   
   \noindent 
    $\nu(\text{CSG}^{(1,0)})  = 1\times 1+2\times(-6)+1\times 20+3\times 9+2\times (-36)+36=0.$
    
    \noindent 
   One finds that in this case:\ 
    $\aa(\text{CSG}^{(1,0)}) = \frac{797}{3840}$, \ 
   $E_{c}(\text{CSG}^{(1,0)}) = \frac{16471}{3840}$. 
   It is possible  to arrange the total  $\aa$-anomaly to vanish   by adding   (1,0)  matter  multiplets (cf. \rf{3.10}) 
   in many ways, none of which seems  particularly special. 
   }

This cancellation  is  the  6d counterpart  of the vanishing of anomalies (and Casimir energy)
 in  the system of 
 $\N=4$ CSG   plus  4  $\N=4$  vector multiplets  in 4d \ci{Fradkin:1985am} where the analog of 
 \rf{4.4}  is 
 \be
\la{4.43}
(\aa,\cc,E_c)(\text{ CSG}^{\N=4}) =  -4\, (\aa,\cc,E_c)(V^{\N=4})\ .  
\ee
 By analogy  with the 4d system  where both a- and c-anomaly coefficients cancel 
  it is  natural to conjecture  also the cancellation of the 
 Weyl-invariant $W_6$ part of the 6d conformal anomaly in \rf{1.1}. 
 This  suggests that \footnote{
 One expects that   the maximal $(2,0)$ supersymmetry implies
  the appearance of a unique
 Weyl invariant  in the conformal anomaly 
  (\ref{1.1}) -- the combination $\mc W_{6}$ in (\ref{444})  which was 
 found in the  free tensor multiplet  case  \cite{Bastianelli:2000hi}
 and also  in  the strong-coupling limit of (2,0) theory as predicted by  the 
 AdS/CFT  \ci{Henningson:1998gx}. Thus   in the (2,0)   case there should  be just one overall 
 $\cc$-anomaly coefficient in \rf{1.1}.
   Then the  same  should   be true   also  for  the (2,0) CSG  (which is equivalent to its 
   1-loop renormalizability).
 Let us  stress  that in 
  the   case of  lower $(1,0)$ supersymmetry this pattern will not apply -- 
    $W_6$   will no longer be 
  universally proportional to  $\mc W_{6}$  --    one will need to consider  two  indepedent  $\cc_i$
  coefficients.
We find  from the expressions in \cite{Bastianelli:2000hi}   that for  the (1,0) scalar  multiplet 
 $S^{(1,0)}$ the $\cc_i$ values are 
 $(\cc_{1},\cc_{2},\cc_{3}) = (-\frac{1}{27},-\frac{1}{540},\frac{1}{180})$ 
 and for the tensor multiplet  $T^{(1,0)}$ they   are  $(-\frac{8}{27},-\frac{11}{135},\frac{1}{45})$ 
  so one has a relation $\cc_1-2 \cc_2 + 6 \cc_3=0$
  (see also  \ci{deBoer:2009pn,Kulaxizi:2009pz}).
  }
 \be\la{445}   \cc(\text{CSG}^{(2,0)}) = -26\,\cc(T^{(2,0)}) =  \frac{13}{144} \ . \ee 
 In this case  the system of (2,0)  CSG coupled to 26 tensor multiplets will  be  
 completely anomaly-free\foot{An
  independent  confurmation of that would be to show the cancellation 
 of axial  anomalies as   was done for the finite 4d  system in \ci{Romer:1985yg}.}
   (in particular, UV finite) 
 and thus  formally consistent   as a quantum theory.
 
 It is interesting  to note  here a possible connection\foot{We thank K. Intriligator
for this  suggestion.}
  to eq.~(3.11) in \cite{Beem:2014kka}  
 $\cc_{(2,0)}/\cc(T^{(2,0)})= c_{\rm 2d}$
 which 
 relates the ratio of c-coefficients  of a  given (2,0)  CFT    and   
   of  the (2,0)  tensor multiplet to  the  central charge  of   some associated 
  2d chiral algebra.  If \rf{445} is true, then in the present case  of (2,0) CSG 
  this ratio  should be  $c_{\rm 2d}= -26$.  It is then natural to  interpret this as 
   central charge or conformal anomaly  of  pure 2d  Einstein gravity (which  has trivial  action and thus is  classically  Weyl invariant):  its anomaly comes just from the ghost determinant
   contributing  the famous  -26  (which can be cancelled   by adding 26 scalar fields as in the bosonic string)  
   \ci{Polyakov:1981rd}. 
  
   Remarkably, a similar relation   exists   for  the   conformal anomaly c-coefficient 
   of  4d  conformal supergravity.
   According to  \ci{Beem:2013sza} in the 
    case of $\N=2$ 4d superconformal theories 
one should have for the associated 2d central charge  
$c_{\rm 2d} = -12  \cc_{\rm 4d}    $
where $ \cc_{\rm 4d}$   is the  4d conformal anomaly c-coefficient (normalised 
to be 1/6 for a vector multiplet). 
This gives 
 $c_{\rm 2d} = -12\times\tfrac{13}{6}=-26$  in the case of  $\N=2$ CSG
   \ci{Fradkin:1981jc}, which should thus  be also  associated to  pure   2d gravity.

 To draw a further     parallel between the  above 4d and 6d results  
   let us recall the relation  of 
 \rf{4.4}  and \rf{4.43}  to quantum properties of KK supermultiplets  appearing in compactification
 of 11d supergravity on AdS$_7 \times $S$^4$  \ci{Beccaria:2014qea}   
 and 10d supergravity on AdS$_5 \times $S$^5$
 \ci{Beccaria:2014xda}  respectively. 
In both cases 
each level of the KK tower is labeled by an integer $p\ge 1$ and contains a set $\Phi^{(p)}$ of 
AdS$_{d+1}$ fields with definite quantum numbers. 
The fields at level $p=1$
represent a (formally  decoupled) 
singleton  -- $(2,0)$ 6d   tensor multiplet   $ T^{(2,0)}$   in the former  case and 
$\N=4$   4d vector multiplet   $V^{\N=4}$    in the latter  case.
The maximal gauged 7d or 5d supergravity fields are found 
 at level $p=2$, while $p>2$ levels contain massive states.
 Eq. (3.4) of \cite{Beccaria:2014qea}  and eq. (6.7) of    \ci{Beccaria:2014xda}  
  give  contributions  to the a-anomaly 
 (proportional to the  coefficient in the 1-loop vacuum partition function in AdS$_{d+1}$)
 of the $\Phi^{(p)}$  supermultiplets  at level $p$  in 6d and 4d cases respectively:
 \be
 \la{4.6}
6d: \ \   \aa(\Phi^{(p)}) = -2\,(6\,p^{2}-6\,p+1)\, \aa(T^{(2,0)}) \ , \qquad 
4d: \ \  \aa(\Phi^{(p)}) = -2 p \, \aa(  V^{\N=4}) \ . 
 \ee
 For $p=1$ we get the singleton multiplet values (taking into account the relation \rf{213} between the  boundary conformal field
 a-coefficient   and the  AdS field  $\aa^+$-coefficient)   while  for $p=2$ we  find the -26  and -4  coefficients 
 in \rf{4.4} and \rf{4.43}.\foot{Curiously, while this  26 coefficient has no apparent  connection to 
 critical dimension of bosonic string (apart from the remark 
 made below eq. \rf{445})
 it originates from the same  type of quadratic polynomial $6\,p^{2}-6\,p+1$   as 
 the ghost  central  charge  of the bosonic string \ci{Polyakov:1981rd} (in the latter case  $p$ plays the role of the spin of the ghost system).}

     \iffa
As was  noticed in  \ci{Beccaria:2014xda}, the  a-anomaly of  4d  $\N=4$ conformal supergravity 
 is  directly  related to the value  of 1-loop   AdS$_5$ vacuum correction in maximal  gauged  5d supergravity.
This remark suggests that 
 the cancellation of a-coefficient in the system  of $\N=4$ CSG + 4 vector multiplets   may be  connected  to the 
 fact that the  total 1-loop   vacuum correction of 10d IIB  supergravity  on AdS$_5 \times $S$^5$ vanishes provided 
 one adds  also the contribution of the (formally   decoupled)   lowest multiplet in  the KK tower -- the boundary singleton 
 represented by  $\N=4$ 4d vector multiplet. 
 In view of the discussion in \ci{Beccaria:2014qea} a  similar interpretation    may  be  given for the 
  cancellation of a-anomaly in  the system of (2,0) CSG + 26 (2,0) tensor multiplets: 
  it may be related to the vanishing of 1-loop vacuum correction in 11d supergravity on AdS$_7 \times $S$^4$
  provided   the contribution of  the singleton (2,0) tensor multiplet is included in the count. 
 \fi

 The  system of (2,0)   conformal  supergravity  coupled to exactly 26    (2,0)  tensor multiplets 
 has  also the following remarkable interpretation. 
 Let  us recall that  starting with $n+5$  (2,0) tensor multiplets  coupled to   (2,0) CSG  background 
 and  spontaneously breaking 
 the dilatation symmetry  by imposing a quadratic constraint \ci{Bergshoeff:1999db,Bergshoeff:1985mz} on  $5\times (n+5)$ tensor 
 multiplet scalars
 one   ends up with  a system of  $n$   tensor    multiplets coupled \cite{Romans:1986er,Riccioni:1997np} 
 to  (2,0)  6d  Poincar\'e supergravity. 
  The remaining $5n$   scalars parametrize  the coset $SO(5,n) \ov SO(5) \times SO(n)$.\foot{
 Explicitly, one imposes 
 $\eta^{IJ} \vp^{ij}_I \vp_{J kl} = M^4 \eta^{ij}_{kl}$,   where $\eta_{IJ}={\rm diag} (-----+...+)$, $I,J=1, ...,5+n$, \  $i,j=1,...,4$ 
   and $M$ is mass scale parameter that determines  the gravitational constant of the 
 Poincar\'e supergravity.
 Here 5 tensor multiplets   are   chosen to be   ghost-like to get 
 the standard physical sign of the Poincar\'e supergravity action. The  number 5 is also directly related to the 
 presence of 5  rank 3  self-dual antisymmetric tensors in the (2,0) CSG spectrum in Table \ref{T2}:
 they couple to the antisymmetric tensor field strength $H_{\m\n\l}$ of the tensor multiplets 
 via $ H_{\mu\nu\rho}  T_{\mu\nu\rho}$ term and (together with  associated   conformal 
 and the S-supersymmetry  fixing)
  this  effectively  eliminates the ghost-like dynamics  of the 5 tensor multiplets. 
  This is a  6d counterpart of  the construction  of $\N=4$  Poincar\'e supergravity coupled to $n$ \   $\N=4$   vector multiplets in 4d by starting 
  with the system of $n+6$\ \ $\N=4$   vector multiplets   in $\N=4$ conformal supergravity background. 
  In this case the 6 ghost-like  vectors  couple to  6 self-dual  tensors  $T_{\m\n}$  of $\N=4$   CSG via 
  $ F_{\mu\nu}  T_{\mu\nu}$, {\em etc.} (for  detailed discussions with  some applications see
   \cite{deRoo:1984gd,Ferrara:2012ui,Carrasco:2013ypa}).}
   Thus,   starting with  (2,0)  CSG coupled to 26=  5 +21  tensor multiplets  and (i)  spontaneously breaking superconformal symmetry  and (ii) decoupling the higher-derivative   terms by considering  a low-energy   limit,
   we end up with chiral (2,0)   6d  Poincar\'e supergravity  coupled to  21   (2,0)    tensor multiplets. The latter theory  is gravitational  anomaly free   (and  also 
   results upon  compactification of type IIB supergravity on K3)  \cite{Townsend:1983xt,Witten:1995em}.
   
 It may be possible to establish a   connection of this   fact  to the a-anomaly  cancellation 
  in  the  (2,0)  CSG  + 26  tensor multiplet system  if  we  assume  that  
   gravitational  anomalies   also  cancel in that theory. As the  chiral  anomalies in the broken and unbroken phases   should match, that   would  suggest 
    (following a  related discussion in the 4d case in section 2.2 of \ci{Carrasco:2013ypa})
   the  cancellation of the gravitational   anomaly also  in  the  corresponding  IR theory
    in the spontaneously broken phase, i.e.  in the   above (2,0)   Poincar\'e supergravity  coupled to  21     tensor multiplets.

%

\section*{Acknowledgments}
We  thank  A. Hanany,  J. Heckman, C. Herzog, K. Intriligator and R. Metsaev 
  for  useful discussions  and correspondence
on related issues.  
The  work of A.A.T. is supported by the ERC Advanced grant No.290456
and also by  the  Russian Science Foundation grant 14-42-00047  associated with Lebedev Institute.


\appendix
\section{Partition functions  of  6d  conformal  fields on $S^6$ 
}
\label{app:s1d6}

{\bf 2-derivative vector $V^{(2)}_{\mu}$} 

Let us start with 2-derivative conformal  6d vector $V^{(2)}_{\mu}$ in \rf{3.1}.
It is 
   a special 
$s=1$  case of a  family of actions for conformal fields  described by 
rank $s$ 
symmetric tensor $\varphi_{\mu_{1}\dots \mu_{s}}$ which  in $d=6$   have   no gauge invariance.
The  corresponding Weyl invariant 
action on curved background was found in \cite{Erdmenger:1997wy} (see also \ci{Beccaria:2015vaa}).
Here we shall use this  action to compute the  a-anomaly  corresponding to $V^{(2)}_{\mu}$   
by direct 6d method  (i.e. from  partition function on $S^6$, see \rf{255}) and  check that it matches the 
value in Table \ref{T1}   found  from \rf{2.5}. 

The Weyl invariant   action for  $V^{(2)}_{\mu}\equiv \vp_\m$  is 
\be
\la{A.1}
S ( V^{(2)}) = \frac{1}{2}\int d^{6}x\sqrt{g}\Big[
\nabla^{\lambda}\varphi^{\mu}\nabla_{\lambda}\varphi_{\mu}-\frac{3}{2}(\nabla^{\lambda}\varphi_{\lambda})^{2}
+\frac{1}{2}R_{\mu\nu}\varphi^{\mu}\varphi^{\nu}+\frac{3}{20}R \varphi^{\mu}\varphi_{\mu}
\Big].
\ee
To diagonalize the kinetic operator, we split $\varphi_{\mu}$ into  transverse and longitudinal parts
\be
\label{A.2}
\varphi_{\mu} = \varphi_{\mu\perp}+\nabla_{\mu}\sigma, \qquad \nabla^{\mu}\varphi_{\mu\perp}=0.
\ee
Specializing to  unit-radius $S^{6}$ (with $R_{\mu\nu} = \frac{R}{6}\,g_{\mu\nu}$, $R =  30$),  we  get 
\be
\la{A.3}
S = \frac{1}{2}\int d^{6}x\sqrt{g}\Big[
\varphi^{\mu}_{\perp}(-\nabla^{2}+7)\varphi_{\mu\perp}-\frac{1}{3}\sigma\,(-\nabla^{2}+6)\, \nabla^{2}\,\,\sigma
\Big].
\ee
Taking into account the Jacobian for the  change of variables in  (\ref{A.2}), we obtain for the partition function
\be
\la{A.4}
Z( V^{(2)}  ) = \Big[\frac{\det\hat\Delta_{0}(0)}{\det\hat\Delta_{1\perp}(7)\,\det\hat\Delta_{0}(6)\det\hat\Delta_{0}(0)}\Big]^{1/2} = 
\Big[\frac{1}{\det\hat\Delta_{1\perp}(7)\,\det\hat\Delta_{0}(6)}\Big]^{1/2},
\ee
where  $\hat\Delta_{1\perp}$  is a  special case   of the operator $\hat\Delta_{s\perp}(M^{2}) = (-\nabla^{2}+M^{2})_{s\perp}$ acting on transverse traceless rank $s$ tensors (with $\hat\Delta_{0}(M^{2}) = -\nabla^{2}+M^{2}$  being scalar Laplacian). 
As a result, we  get   from \rf{255} 
\be
\label{A.5}
\aa( V^{(2)}  ) = -\frac{1}{96}\Big(B_{6}[\hat\Delta_{1\perp}(7)]+B_{6}[\hat\Delta_{0}(6)]\Big)\ ,
\ee
where  $B_{6}$ are  the corresponding Seeley coefficients.  The  general expression  for 
$B_{6}[\hat\Delta_{s\perp}(M^{2})] $  (applicable also to $s=0$ case) 
was found   in \cite{Tseytlin:2013fca} 
\be
\la{A.6}
\begin{split}
B_{6}[\hat\Delta_{s\perp}(M^{2})] &= \frac{(s+1)(s+2)(2s+3)}{453600}\Big[-210\,M^{6}-315\,M^{4}(s^{2}+s-10) \\
& +630\,M^{2}(s+6)(s^{2}+2s-4) \\
&+22780-18774 s -19488 s^{2}-4515 s^{3}-315 s^{4}\Big].
\end{split}
\ee
Computing the two terms in (\ref{A.5}), we finally get
\be
\la{A.7}
\aa( V^{(2)}  ) = -\frac{1}{96}\Big(\frac{67}{3780}+\frac{1}{756}\Big) = -\frac{1}{7!} \ . 
\ee
This is in agreement with the $V^{(2)}$ entry in Table~\ref{T1} derived  from \rf{2.5}, i.e. 
using AdS$_7$-based method.

\

 {\bf 4-derivative vector $V^{(4)}_{\mu}$} 

The  4-derivative     $V^{(4)}_\m\equiv A_\m$  vector field in \rf{3.1}   with  gauge-invariant  action 
 is the $s=1$   member of the conformal higher  spin (CHS)  family in $d=6$ \ci{Tseytlin:2013fca}. 
 Considering  generic curved background  and assuming  that  under Weyl  transformations $g'_{\m\n} = e^{2 \rho} g_{\m\n}$    the gauge field 
 is not transforming  
 $A'_\m = A_\m$  we find  that the corresponding  Weyl-invariant action  is (here $ F_{\m\n} = \del_\m A_\n - \del_\n A_\m$)\foot{We use that 
   for $J= {1 \ov 2( d-1) }R, \ K_{\m\n} = {1 \ov d-2} ( R_{\m\n} - J g_{\m\n})$ in $d$ dimensions  one has 
 $\delta J=- 2 \r J -  \nabla^2 \r, \ \ \delta K_{\m\n} =- \nabla_\m \nabla_\n  \r$. One can also add with an arbitrary coefficient 
  a  term with the Weyl tensor  
 $\int d^{6}x\sqrt{g}   C_{\m\k\n\l } F^{\m\n} F^{\k\l } $ which is separately Weyl-invariant.}
 \be
\la{A.101}
S ( V^{(4)}) = \int d^{6}x\sqrt{g}\Big[
\nabla_{\lambda}   F^{\l}_{\  \ \m}  \nabla_{\nu}   F^{\n \m}  
- ( R_{\m\n} -   { 1 \ov 5}   R g_{\m\n})  F^{\m\l}  F^\n_{\ \ \l} \Big]
 \ . \ee
Integrating by parts and using $\del_{[\m} F_{\n\l]}=0$   this  action can be written also as 
\be
\la{A.1111}
S ( V^{(4)}) =\ha  \int d^{6}x\sqrt{g}\  F^{\m\n}  \Big( - \nabla^2  F_{\m\n}  + { 2 \ov 5}   R   F_{\m\n }   - 2 R_{\m\k\n\l }F^{\k\l } \Big)
\   . \ee
Note  that \rf{A.101}  may be viewed  as a 6d Weyl-invariant action  for a 
  zero Weyl weight  2-form field $B_{\m\n}=F_{\m\n}$   only assuming it satisfies 
  the Bianchi  constraint, i.e. is longitidinal. 
An alternative Weyl invariant action that depends 
only on  the  transverse part of $B$ is  the familiar one  $ \int d^{6}x\sqrt{g}\   H^{\m\n\l} H_{\m\n\l}$ where $H=dB$ 
(cf. \ci{Erdmenger:1997gy}). 

Specifying to $S^6$   with $R_{\m\k\n\l }= g_{\m\n} g_{\k\l} - g_{\m\l} g_{\k\n}$ we   get  from either \rf{A.101} or \rf{A.1111}  
with $ F_{\m\n} = \del_\m A_\n - \del_\n A_\m$
(cf. \rf{A.3}) 
\be
\la{A.311}
S ( V^{(4)}) = \int d^{6}x\sqrt{g}\ 
A^{\mu}_{\perp}(-\nabla^{2}+7)(-\nabla^{2}+5)A_{\mu\, \perp} \ . 
\ee
 Taking into account the Jacobian of the change of variables $A_\m\to A_{\mu\perp} + \nabla_\m \sigma  $ 
    one finds then the 
     following partition function  (note that $ \nabla_{\lambda}   F^{\l}_{\  \ \m}  = (-\nabla^{2}+5)A_{\mu\, \perp}$)
(cf.  \rf{A.4})
\be
\la{A.481}
Z( V^{(4)}  ) = 
\Big[\frac{ \det\hat\Delta_{0}(0)} {\det\hat\Delta_{1\perp}(7)\,\det\hat\Delta_{1\perp}(5)}\Big]^{1/2} \ .
\ee
This     agrees with the $s=1$ case   of  
the general  expression for the 6d partion function  of the   CHS field  in  \cite{Tseytlin:2013fca}. 

Using \rf{A.6} that gives (cf. \rf{A.7}) 
\be
\la{A.71}
\aa( V^{(4)}  ) = -\frac{1}{96}\Big(\frac{67}{3780} -\frac{1403}{3780}- \frac{1139}{3780}  \Big) = \frac{1}{7!} { 275 \ov 8}\ ,
\ee
in agreement with the $V^{(4)}$  value  in Table~\ref{T1}. 
Here the middle entry is that of  the operator $\hat\Delta_{1\perp}(5)$.
The latter   appears also 
in the  $S^6$ partition   function of the 
standard (non-conformal)  6d Maxwell  vector   
(here  $- g_{\m\n} \nabla^2 + R_{\m\n} \to  g_{\m\n}  ( - \nabla^2 + 5)$, cf. \rf{A.4}, \rf{A.481}) 
\be
\la{A8}
Z({\rm  6d\,  Maxwell} ) = 
\Big[\frac{ \det\hat\Delta_{0}(0)} {\det\hat\Delta_{1\perp}(5)}\Big]^{1/2} \ . 
\ee
The difference between \rf{A8}  and \rf{A.481} is thus in the  contribution of  the operator 
$\hat\Delta_{1\perp}(7)$  which is  the same as 
the transverse part of the $V^{(2)}$ operator  in \rf{A.3}.
For (\ref{A8}),  the  coefficient of  logarithmically divergent part of $\log Z$ in \rf{255} 
(which  here does not  have  the standard  conformal anomaly interpretation so we will call it  $"\aa" $) is\foot{
Then for (1,0) Maxwell multiplet containing in addition two  MW fermions $\psi$ in Table \ref{T1}   we get
$"\aa" ({\rm (1,0) \, Maxwell}) =   \frac{1}{7!}  ( { 1271 \ov 36} - { 191 \ov 144} ) =      \frac{1}{7!} { 1631 \ov 48}$.}
\be
\la{A.711}
"\aa" ({\rm  6d\,  Maxwell} )= -\frac{1}{96}\Big(-\frac{1403}{3780}- \frac{1139}{3780}  \Big) =  \frac{1}{7!} { 1271 \ov 36}\ .
\ee
In  6  dimensions  a  gauge-invariant vector $A_\m$ is dual  to rank 3 antisymmetric tensor $C_{\m\n\l}$, e.g., 
   the Maxwell   action $F_{\m\n} F^{\m\n}$  is   dual  to $H_{\m\n\l\r} H^{\m\n\l\r}$  where $H_4= dC_3$.
Starting  with flat space case  and  considering  instead  the  conformally   invariant 4-derivative $ V^{(4)}$   Lagrangian  
$L= F_{\m\n} \del^2 F^{\m\n}$  we get as  its dual a non-local 
$\td L =- { 1 \ov 4}  H^{\m\n\l\r} \del^{-2} H_{\m\n\l\r} =  C^{\m\n\l}_ \perp  C_{\m\n\l \, \perp}$, where $\del^\m C_{\m\n\l\, \perp } =0$.
The corresponding  action is conformal  and may be viewed as the analog of the conformal Schwinger action 
 $\int d^2 x (A_{\m \, \perp})^2$ for a  vector in 2 dimensions. 
 
 Dualizing   the  curved space   action \rf{A.101} or \rf{A.1111} by first adding $\epsilon^{\m\n\l\r\k\delta}  F_{\m\n}  H_{ \l\r\k\delta}$
 and then  integrating out $F_{\m\n}$ one ends up  with a  non-local action for $H_{ \l\r\k\delta}$.
For example, for $S^6$ background when the integrand in \rf{A.1111}  becomes $ 
F^{\m\n}  \big( - \nabla^2  +10 \big) F_{\m\n } $  we get\foot{
Note that $\nabla^\m H_{\m\n\l\r}  = \big( - \nabla^2  +9\big)C_{\n\l\r \, \perp}$.  
In general, for Hodge-deRham operator acting on a $p$-form in $d$ dimensions one has 
(see, e.g.,  \ci{Elizalde:1996nb}): $\hat \Delta = -\nabla^2 + p (d-p)$. 
To integrate over the   2-form  $F$ one needs to split $F= F_\perp + d A$
which brings in the Jacobian $[\det\hat  \Delta_{1\perp}(5)]^{-1/2}$
and to notice that $C_3$ couples only to $F_\perp$.
 } 
 \def \na {\nabla}
\be \la{A.99}
\td L =- { 1 \ov 4}  H^{\m\n\l\r}  \big( - \nabla^2  +10 \big)^{-1}_\perp    H_{\m\n\l\r} = 
C^{\m\n\l}_ \perp \big( - \nabla^2  +9\big)_\perp  \big( - \nabla^2  +11 \big)^{-1}_\perp   C_{\m\n\l \, \perp} \ . 
\ee
To show this one notes that the Lagrangian  may be written as 
$-C^{\m\k\l} \nabla^\r \frac{1}{-\nabla^2+10} \nabla_\r C_{\m\k\l}-3 C^{\m\k\l} \na^\r \frac{1}{-\nabla^2+10} \na_\l C_{\m\k\r}.
$ Then one  may expand $\tfrac{1}{-\nabla^2+10} = \tfrac{1}{10} (1+\tfrac{1}{10} \nabla^2+\tfrac{1}{100}\nabla^4+\dots)$ and commute 
 covariant derivatives using that only the transverse part of $C_3$ should 
  contribute.\foot{Explicitly,  $-C^{\m\k\l\, }_\perp \nabla^\r (\na^2)^n  \nabla_\r C_{\m\k\l\, \perp}-3 C^{\m\k\l\,}_ \perp \na^\r  (\na^2)^n \na_\l C_{\m\k\r\, \perp} = C^{\m\k\l\, \perp }  P_n (\na^2) C_{\m\k\l\, \perp } 
$
where $P_n$ is a polynomial in $\nabla^2$   which can be shown to be  $P_n = \frac{1}{10^{n+1}} (9-\nabla^2)  ( \nabla^2-1)^n$. }

The dual   $C_3$   theory is thus defined by the path integral
 $Z(C)= \int [d C_3]\   G \, e^{-\td S(C_3)}$  with the   gauge-invariant    but non-local action 
\rf{A.99}  and  $G= [\det\hat  \Delta_{F}(10)]^{-1/2}$    (with $\hat  \Delta_{F}(M^2) \equiv - \na^2 + M^2$ acting on a rank 2  antisymmetric  tensor)   being  the  factor coming from integrating out $F$ in the 1st order action
$F_2  \Delta_{F}(10)  F_2 +   \epsilon_6   F_2  H_4$.
  Equivalently,  we may write 
$G=[\det\hat  \Delta_{F \perp}(10)\, 
\det \hat \Delta_{1\perp}(7)]^{-1/2} $. 
This unfamiliar  factor (irrelevant at the classical level) is important for quantum equivalence of the two dual theories. 
One can   show that the partition function resulting from integration over $C_3$  is  indeed equivalent to \rf{A.481}.
Splitting  $C_3=C_{3\perp} +dB_2$ where $B_2$
is a 2-form that can be  also  decomposed as $B_2=B_{2\perp}+dA$, etc., 
we get the   Jacobian    which is 
the inverse of the standard 2-form partition function in \rf{A91}, i.e. 
$J= 1/Z(B) = \big[{\det\hat\Delta_{B\perp}(8)\,\det\hat\Delta_{0}(0)}
\big]^{1/2} \big[{\det\hat\Delta_{1\perp}(5)}\big]^{-1/2}$.
Integrating out $C_{3\perp}$ gives 
$\big[ \det  \hat \Delta_{C\perp}  (9)  \big]^{-1/2} \big[  \hat \Delta_{C\perp}  (11)\big]^{1/2}.$
Now using  the relation  between $S^6$  determinants  of 2-form and 3-form operators  \ci{Elizalde:1996nb} 
$
\det(-\na ^{2}+m^2)_{B\perp} = \det(-\na^{2}+m^2+1)_{C\perp} 
$
one  finds that $Z(C)$ is equivalent to $Z(V^{(4)})$ in \rf{A.481}. 

A potential   interest in considering  the action for $C_3$ like   $C_{3\perp} C_{3\perp} + ...$ as a dual alternative 
to the conformal 4-derivative vector theory 
is that  a similar term appeared  in the context of (1,0) superconformal models with  
non-abelian tensor fields  \ci{Samtleben:2011fj,Samtleben:2012mi}. 
For example, a conformal model $L= ( C_3 + d B_2)^2   + \phi ( B_2 + d A)^2  + ...$
in the phase where  $\phi$ has trivial  expectation value  will  have  its conformal anomaly   determined  by the first term 
  which has an effective  gauge invariance, i.e. by  $(C_{3 \perp})^2  + (d B_2')^2$. 
  The contribution  of $(C_{3 \perp})^2$ to the conformal anomaly will  be coming just from the measure  or ghost factors  and will thus 
 be like that of a non-unitary   theory.

\

\

 {\bf Antisymmetric tensor $T_{\m\n}$} 

The $S^6$ partition function for  antisymmetric tensor $T_{\m\n}$  with the standard gauge-invariant  Lagrangian 
$H_{\m\n\l} H^{\m\n\l}, \  H= d T$  (we relax self-duality condition here)    can   be  found, {\em e.g.},  
 from the general curved space 
expression in \ci{Fradkin:1982kf,Bastianelli:2000hi}
\be \la{A91}
Z(T) = \Big[
\frac{(\det\hat\Delta_{1\perp}(5))^2}{\det\hat\Delta_{T}(8)\,(\det\hat\Delta_{0}(0))^3}
\Big]^{1/2}
=
 \Big[
\frac{\det\hat\Delta_{1\perp}(5)}{\det\hat\Delta_{T\perp}(8)\,\det\hat\Delta_{0}(0)}
\Big]^{1/2},
\ee
where $\hat\Delta_{T} T_{\m\n} = ( - \nabla^2 + M^2) T_{\m\n}$   and $\hat\Delta_{T\perp}$ 
acts on transverse antisymmetric 2-tensors.\foot{Here the first  equality  is found in the usual covariant Feynman gauge while to derive  the second one 
we  should set  $T_{\m\n} = T_{ \m\n\, \perp} + B_{\m\n}\ , \  \  B_{\m\n}=  \nabla_\m Q_\nu - \nabla_\n Q_\m$,    and account for 
the  Jacobian $J_1= \big[ \det\hat\Delta_{1\perp}{(5)}\big]^{1/2}$ of this   change   of variables. 
Note that     without  gauge fixing the   classical Lagrangian  is $H_{\m\n\l} H^{\m\n\l} \sim  T_\perp  \hat\Delta_{T\perp}(8) T_\perp $
and the  remaining factors 
 in the second form of  \rf{A91} come from the  Jacobian  $J_1$ 
and the fact that one is  to divide  out the full  gauge group volume 
$[d Q_\m] = d Q_{\m\perp}\, d \alpha\,  [ \det\hat\Delta_{0}{(0)}\big]^{1/2}$
(here we set $Q_\m = Q_{\m\perp} + \del_\m \alpha$).
Note also  that 
on a general curved background 
$( \hat\Delta_{T})^{\l\r}_{\m\n} = - ( \nabla^2)^{\l\r}_{\m\n} +  2 R^{[\l}_{[\m} \delta^{\r]}_{\n]}   - R^{\l\r}_{\ \ \m\n} $ so that 
for a unit-radis  $S^6$ where $R_{\l\r\m\n} = g_{\l\m} g_{\r\n} - g_{\l\n} g_{\r\m}$
we  thus   get  $( \hat\Delta_{T})^{\l\r}_{\m\n} =  ( -\nabla^2 + 8)^{\l\r}_{\m\n}  $.
The action  before gauge fixing depends only on $T_{\m\n\, \perp}$   while the standard gauge-fixing term $(\nabla^\m T_{\m\n})^2$ 
gives $(\nabla^\m C_{\m\n})^2$   or $\big[ (-\nabla^2 + 5) C_{\m\, \perp}  \big]^2$.
} 
Note that  the partition function of $T$-field  with  the contribution from its action $\det\hat\Delta_{T\perp}(8)$  omitted is just the 
inverse of the  6d Maxwell  partition function \rf{A8}.

\


{\bf Relation to conformal group representations }

Comparing the structure  of the  above   6d partition functions on $S^6$  to the conformal   group representation content of the respective 
fields  in Table \ref{T1}   we conclude   that the correspondence is not  immediately straightforward. 
In general,  given a field in some representation $(\Delta\equiv \De_+; h_1,h_2,h_3)$  it   
will correspond to $Z= D^{-1/2}$  where $D$ is some product of determinants of 2nd order Laplacains on $S^6$. 
This is, indeed, expected   for higher derivative  conformal  scalar (GJMS)  operators. 
 For totally symmetric traceless   conformal tensors  the general rule 
appears to be 
\be
\label{A.11}
(\Delta; s,0,0) \longrightarrow  \prod_{k=1}^{\Delta-3}\prod_{s'=0}^{s}\det\hat\Delta_{s'\,\perp}(M^{2}_{s',k-2})\ ,\qquad 
\qquad  M^{2}_{s',k-2} = 6+s'-k(k-1) \ , 
\ee
where   $M^{2}_{k,n} =k-(n-1)(n+d-2) $ or  $   k-(s-1)(s+4)$ in $d=6$.
One can check that the corresponding a-coefficient computed via 7d relation \rf{2.5} is the same as  found 
using 6d approach based on \rf{A.6}, {\em i.e.} 
\be\la{a15}
\aa(\Delta; s,0,0) = -\frac{1}{96}\,\sum_{k=1}^{\Delta-3}\sum_{s'=0}^{s}
B_{6}[\hat\Delta_{s'\,\perp}(M^{2}_{s',k-2})].
\ee
The case of $s=0$   corresponds  to scalar GJMS operators (see, e.g., \cite{Graham:2007,
Manvelyan:2006bk,Juhl:2011aa})
 where  (\ref{A.11}) reproduces  the known relations, {\em e.g.}, 
\be \la{a16} 
(4,0,0,0)\longrightarrow \det\hat\Delta_{0}(6) \ , \ \ \ \ \ \ \ 
(6,0,0,0)\longrightarrow \det\hat\Delta_{0}(6)\,\det\hat\Delta_{0}(4)\,\det\hat\Delta_{0}(0).
\ee
For $s=1$ some relevant  special cases are 
\be
\begin{split}\la{a17} 
 (4; 1,0,0) & \longrightarrow \det\hat\Delta_{1\perp}(7)\,\det\hat\Delta_{0}(6)\ , \\
 (5; 1,0,0) & \longrightarrow \det\hat\Delta_{1\perp}(7)\, \det\hat\Delta_{1\perp}(5)
\,\det\hat\Delta_{0}(6)\,\det\hat\Delta_{0}(4) \ .
\end{split}
\ee
Since,  according to Table \ref{T1},    the vector $V^{(2)}$  corresponds to $(4;1,0,0)$
and $V^{(4)}$  to $(5;1,0,0)-(6;0,0,0)$ representations,  we see that \rf{a17},\rf{a16} are indeed consistent with 
the  partition functions in \rf{A.4} and \rf{A.481}. 

\iffa 
Using these one can re-obtain the partition function of $V^{(2)}$ and $V^{(4)}$ by simply putting together
the pieces associated with the relevant representations. For $V^{(2)}$, one has trivially
\be
Z(V^{(2)}) = Z[(4;1,0,0)] = \bigg[
\frac{1}{\det\hat\Delta_{1\perp}(7)\,\det\hat\Delta_{0}(6)}
\bigg]^{1/2},
\ee
in agreement with (\ref{A.4}). For $V^{(4)}$, one simply combines physical and ghost fields and gets
\be
\begin{split}
Z(V^{(4)}) &= Z[(5;1,0,0)-(6; 0,0,0)] = \\
&=\bigg[
\frac{\det\hat\Delta_{0}(6)\,\det\hat\Delta_{0}(4)\,\det\hat\Delta_{0}(0)}{\det\hat\Delta_{1\perp}(7)\, \det\hat\Delta_{1\perp}(5)
\,\det\hat\Delta_{0}(6)\,\det\hat\Delta_{0}(4)}
\bigg]^{1/2} = \bigg[
\frac{\det\hat\Delta_{0}(0)}{\det\hat\Delta_{1\perp}(7)\, \det\hat\Delta_{1\perp}(5)}
\bigg]^{1/2},
\end{split}
\ee
in agreement with (\ref{A.481}). 
\fi

For $s=2$   the important  special case is the conformal graviton in Table \ref{T2}. 
Here  one finds  from \rf{A.11}
\be
Z(h_{\mu\nu}) = Z[(6;2,0,0)-(7; 1,0,0)] 
=
 \Big[
\frac{\det\hat\Delta_{1\perp}(-5)\,
 \det\hat\Delta_{0}(-6)}{ \det\hat\Delta_{2\perp}(8)\,\det\hat\Delta_{2\perp}(6)\,\det\hat\Delta_{2\perp}(2)}
\Big]^{1/2} \ , 
\ee
which is in agreement with the $s=2$ CHS   expression  in 
 \cite{Tseytlin:2013fca}  (see also    \ci{Pang:2012rd}).  

Similarly, in  the antisymmetric tensor case $ T=  (4; 1,1,0) - (5; 1,0,0) + (6; 0,0,0) $   (cf.  Table \ref{T1}) 
 eq.\rf{A91} together with \rf{a16},\rf{a17} 
 imply  the  following correspondence
\be
\label{A.21}
(4; 1,1,0)\longrightarrow \det\hat\Delta_{T \perp}(8)\,\det\hat\Delta_{1\perp}(7).
\ee

\iffa 
Finally, let us examine the case of a (non-selfdual) tensor. This is interesting because it involves a 
 {\em highest spin}
representation that is not of the form $(\Delta; s, 0, 0)$.
Explicit analysis leads to the partition function \cite{Bastianelli:2000hi} \footnote{
Here, count of degrees of freedom can be checked as follows. $\det\hat\Delta_{T\perp}$
corresponds to  $6 \times\frac{5}{2}=15$ degrees of freedom. We subtract the 6 conditions
$ \partial^{\mu}T_{\mu\nu}=0$ and adds back one because 
of the algebraic identity $\partial^{\mu}\partial^{\nu}T_{\mu\nu}=0$. 
So $\nu(T) = 10-5+1=6$ correct for a non-selfdual tensor in 6d.
} \red{[is it correct to cite Bastianelli in Feynman gauge to write (\ref{A.20}) ? }
\be
\label{A.20}
Z(T) = \bigg[
\frac{\det\hat\Delta_{1\perp}(5)}{\det\hat\Delta_{T\perp}(M_{T}^{2})\,\det\hat\Delta_{0}(M_{S}^{2})}
\bigg]^{1/2},
\ee
where $\hat\Delta_{T\perp}$ acts on transverse antisymmetric 2-tensors. We do not have $B_{6}$ for such a tensor,
but we can use the known part of the dictionary to determine the 
determinants associated with the representation $(4; 1,1,0)$. Assuming $M_{S}^{2}=0$, we find
\be
\label{A.21}
(4; 1,1,0)\longrightarrow \det\hat\Delta_{T \perp}(M_{T}^{2})\,\det\hat\Delta_{1\perp}(7).
\ee
Indeed, with this correspondence, 
\be
\begin{split}
Z(T) &= Z[(4;1,1,0)-(5,1,0,0)+(6,0,0,0)] = \\
&=\bigg[
\frac{ \det\hat\Delta_{1\perp}(7)\, \det\hat\Delta_{1\perp}(5)
\,\det\hat\Delta_{0}(6)\,\det\hat\Delta_{0}(4)}{\det\hat\Delta_{T \perp}(M_{T}^{2})\,\det\hat\Delta_{1\perp}(7)\times
\det\hat\Delta_{0}(6)\,\det\hat\Delta_{0}(4)\,\det\hat\Delta_{0}(0)
}
\bigg]^{1/2} \\
&= \bigg[
\frac{\det\hat\Delta_{1\perp}(5)}{\det\hat\Delta_{T \perp}(M_{T}^{2})\,\det\hat\Delta_{0}(0)
}
\bigg]^{1/2} ,
\end{split}
\ee
in agreement with (\ref{A.20}). \red{If we had 7 in numerator of (\ref{A.20}), then we would have 5 in second term
in (\ref{A.21}). However, according to Bastianelli and your messages I would say that 5 in numerator of (\ref{A.20}) should be ok as well as $M_{S}^{2}=0$.}. The correspondence (\ref{A.21}) may be interpreted as the combination of 
the pure $T_{\perp}$ contribution plus its lower spin 1 $C_{\mu}$ component appearing in the decomposition $T_{\mu\nu} = T_{\mu\nu\perp}+
\partial_{[\mu}C_{\nu]}$.
 \fi 


\bibliography{HigherSpin6d-Biblio}

\providecommand{\href}[2]{#2}\begingroup\raggedright\begin{thebibliography}{10}

\bibitem{Cordova:2015vwa}
C.~Cordova, T.~T. Dumitrescu, and X.~Yin, {\it {Higher Derivative Terms,
  Toroidal Compactification, and Weyl Anomalies in Six-Dimensional (2,0)
  Theories}},  \href{http://arxiv.org/abs/1505.03850}{{\tt arXiv:1505.03850}}.

\bibitem{Cordova:2015fha}
C.~Cordova, T.~T. Dumitrescu, and K.~Intriligator, {\it {Anomalies,
  Renormalization Group Flows, and the a-Theorem in Six-Dimensional (1,0)
  Theories}},  \href{http://arxiv.org/abs/1506.03807}{{\tt arXiv:1506.03807}}.

\bibitem{HH}
J.~Heckman and C.~Herzog, {\it {a Conformal Anomaly for 6d SCFT's}},
  \href{http://arxiv.org/abs/private communication}{{\tt private
  communication}}.

\bibitem{Bonora:1985cq}
L.~Bonora, P.~Pasti, and M.~Bregola, {\it {Weyl cocycles}},  {\em
  Class.Quant.Grav.} {\bf 3} (1986) 635.

\bibitem{Deser:1993yx}
S.~Deser and A.~Schwimmer, {\it {Geometric classification of conformal
  anomalies in arbitrary dimensions}},  {\em Phys.Lett.} {\bf B309} (1993)
  279--284, [\href{http://arxiv.org/abs/hep-th/9302047}{{\tt hep-th/9302047}}].

\bibitem{Bastianelli:2000hi}
F.~Bastianelli, S.~Frolov, and A.~A. Tseytlin, {\it {Conformal anomaly of (2,0)
  tensor multiplet in six-dimensions and AdS / CFT correspondence}},  {\em
  JHEP} {\bf 0002} (2000) 013, [\href{http://arxiv.org/abs/hep-th/0001041}{{\tt
  hep-th/0001041}}].

\bibitem{Tseytlin:2013fca}
A.~A. Tseytlin, {\it {Weyl anomaly of conformal higher spins on six-sphere}},
  {\em Nucl.Phys.} {\bf B877} (2013) 632--646,
  [\href{http://arxiv.org/abs/1310.1795}{{\tt arXiv:1310.1795}}].

\bibitem{Giombi:2013yva}
S.~Giombi, I.~R. Klebanov, S.~S. Pufu, B.~R. Safdi, and G.~Tarnopolsky, {\it
  {AdS Description of Induced Higher-Spin Gauge Theory}},  {\em JHEP} {\bf
  1310} (2013) 016, [\href{http://arxiv.org/abs/1306.5242}{{\tt
  arXiv:1306.5242}}].

\bibitem{Giombi:2014iua}
S.~Giombi, I.~R. Klebanov, and B.~R. Safdi, {\it {Higher Spin
  AdS$_{d+1}$/CFT$_d$ at One Loop}},  {\em Phys.Rev.} {\bf D89} (2014) 084004,
  [\href{http://arxiv.org/abs/1401.0825}{{\tt arXiv:1401.0825}}].

\bibitem{Beccaria:2014xda}
M.~Beccaria and A.~A. Tseytlin, {\it {Higher spins in AdS$_{5}$ at one loop:
  vacuum energy, boundary conformal anomalies and AdS/CFT}},  {\em JHEP} {\bf
  1411} (2014) 114, [\href{http://arxiv.org/abs/1410.3273}{{\tt
  arXiv:1410.3273}}].

\bibitem{Beccaria:2014qea}
M.~Beccaria, G.~Macorini, and A.~A. Tseytlin, {\it {Supergravity one-loop
  corrections on AdS$_7$ and AdS$_3$, higher spins and AdS/CFT}},  {\em
  Nucl.Phys.} {\bf B892} (2015) 211--238,
  [\href{http://arxiv.org/abs/1412.0489}{{\tt arXiv:1412.0489}}].

\bibitem{Ivanov:2005qf}
E.~Ivanov, A.~V. Smilga, and B.~Zupnik, {\it {Renormalizable supersymmetric
  gauge theory in six dimensions}},  {\em Nucl.Phys.} {\bf B726} (2005)
  131--148, [\href{http://arxiv.org/abs/hep-th/0505082}{{\tt hep-th/0505082}}].

\bibitem{Ivanov:2005kz}
E.~Ivanov and A.~V. Smilga, {\it {Conformal properties of hypermultiplet
  actions in six dimensions}},  {\em Phys.Lett.} {\bf B637} (2006) 374--381,
  [\href{http://arxiv.org/abs/hep-th/0510273}{{\tt hep-th/0510273}}].

\bibitem{Smilga:2006ax}
A.~V. Smilga, {\it {Chiral anomalies in higher-derivative supersymmetric 6D
  theories}},  {\em Phys.Lett.} {\bf B647} (2007) 298--304,
  [\href{http://arxiv.org/abs/hep-th/0606139}{{\tt hep-th/0606139}}].

\bibitem{Fradkin:1985am}
E.~S. Fradkin and A.~A. Tseytlin, {\it {Conformal supergravity}},  {\em
  Phys.Rept.} {\bf 119} (1985) 233--362.

\bibitem{Erdmenger:1997wy}
J.~Erdmenger and H.~Osborn, {\it {Conformally covariant differential operators:
  Symmetric tensor fields}},  {\em Class.Quant.Grav.} {\bf 15} (1998) 273--280,
  [\href{http://arxiv.org/abs/gr-qc/9708040}{{\tt gr-qc/9708040}}].

\bibitem{ElShowk:2011gz}
S.~El-Showk, Y.~Nakayama, and S.~Rychkov, {\it {What Maxwell Theory in $D\neq
  4$ teaches us about scale and conformal invariance}},  {\em Nucl.Phys.} {\bf
  B848} (2011) 578--593, [\href{http://arxiv.org/abs/1101.5385}{{\tt
  arXiv:1101.5385}}].

\bibitem{Samtleben:2011fj}
H.~Samtleben, E.~Sezgin, and R.~Wimmer, {\it {(1,0) superconformal models in
  six dimensions}},  {\em JHEP} {\bf 1112} (2011) 062,
  [\href{http://arxiv.org/abs/1108.4060}{{\tt arXiv:1108.4060}}].

\bibitem{Fradkin:1983tg}
E.~S. Fradkin and A.~A. Tseytlin, {\it {Conformal Anomaly in Weyl Theory and
  Anomaly Free Superconformal Theories}},  {\em Phys.Lett.} {\bf B134} (1984)
  187.

\bibitem{Bergshoeff:1999db}
E.~Bergshoeff, E.~Sezgin, and A.~Van~Proeyen, {\it {(2,0) tensor multiplets and
  conformal supergravity in D = 6}},  {\em Class.Quant.Grav.} {\bf 16} (1999)
  3193--3206, [\href{http://arxiv.org/abs/hep-th/9904085}{{\tt
  hep-th/9904085}}].

\bibitem{Townsend:1983xt}
P.~Townsend, {\it {A New Anomaly Free Chiral Supergravity Theory From
  Compactification on K3}},  {\em Phys.Lett.} {\bf B139} (1984) 283.

\bibitem{Witten:1995em}
E.~Witten, {\it {Five-branes and M theory on an orbifold}},  {\em Nucl.Phys.}
  {\bf B463} (1996) 383--397, [\href{http://arxiv.org/abs/hep-th/9512219}{{\tt
  hep-th/9512219}}].

\bibitem{Metsaev:1995re}
R.~R. Metsaev, {\it {Massless mixed symmetry bosonic free fields in
  d-dimensional anti-de Sitter space-time}},  {\em Phys.Lett.} {\bf B354}
  (1995) 78--84.

\bibitem{Dolan:2005wy}
F.~Dolan, {\it {Character formulae and partition functions in higher
  dimensional conformal field theory}},  {\em J.Math.Phys.} {\bf 47} (2006)
  062303, [\href{http://arxiv.org/abs/hep-th/0508031}{{\tt hep-th/0508031}}].

\bibitem{Shaynkman:2004vu}
O.~Shaynkman, I.~Y. Tipunin, and M.~Vasiliev, {\it {Unfolded form of conformal
  equations in M dimensions and $\mathfrak{o}(M + 2)$ modules}},  {\em
  Rev.Math.Phys.} {\bf 18} (2006) 823--886,
  [\href{http://arxiv.org/abs/hep-th/0401086}{{\tt hep-th/0401086}}].

\bibitem{Bekaert:2013zya}
X.~Bekaert and M.~Grigoriev, {\it {Higher order singletons, partially massless
  fields and their boundary values in the ambient approach}},  {\em Nucl.Phys.}
  {\bf B876} (2013) 667--714, [\href{http://arxiv.org/abs/1305.0162}{{\tt
  arXiv:1305.0162}}].

\bibitem{Barnich:2015tma}
G.~Barnich, X.~Bekaert, and M.~Grigoriev, {\it {Notes on conformal invariance
  of gauge fields}},  \href{http://arxiv.org/abs/1506.00595}{{\tt
  arXiv:1506.00595}}.

\bibitem{Barvinsky:2005ms}
A.~Barvinsky and D.~Nesterov, {\it {Quantum effective action in spacetimes with
  branes and boundaries}},  {\em Phys.Rev.} {\bf D73} (2006) 066012,
  [\href{http://arxiv.org/abs/hep-th/0512291}{{\tt hep-th/0512291}}].

\bibitem{Barvinsky:2014kta}
A.~Barvinsky, {\it {Holography beyond conformal invariance and AdS isometry?}},
   {\em J.Exp.Theor.Phys.} {\bf 120} (2015), no.~3 449--459,
  [\href{http://arxiv.org/abs/1410.6316}{{\tt arXiv:1410.6316}}].

\bibitem{Gubser:2002vv}
S.~S. Gubser and I.~R. Klebanov, {\it {A Universal result on central charges in
  the presence of double trace deformations}},  {\em Nucl.Phys.} {\bf B656}
  (2003) 23--36, [\href{http://arxiv.org/abs/hep-th/0212138}{{\tt
  hep-th/0212138}}].

\bibitem{Diaz:2007an}
D.~E. Diaz and H.~Dorn, {\it {Partition functions and double-trace deformations
  in AdS/CFT}},  {\em JHEP} {\bf 0705} (2007) 046,
  [\href{http://arxiv.org/abs/hep-th/0702163}{{\tt hep-th/0702163}}].

\bibitem{Diaz:2008hy}
D.~E. Diaz, {\it {Polyakov formulas for GJMS operators from AdS/CFT}},  {\em
  JHEP} {\bf 0807} (2008) 103, [\href{http://arxiv.org/abs/0803.0571}{{\tt
  arXiv:0803.0571}}].

\bibitem{Tseytlin:2013jya}
A.~A. Tseytlin, {\it {On partition function and Weyl anomaly of conformal
  higher spin fields}},  {\em Nucl.Phys.} {\bf B877} (2013) 598--631,
  [\href{http://arxiv.org/abs/1309.0785}{{\tt arXiv:1309.0785}}].

\bibitem{Beccaria:2014jxa}
M.~Beccaria, X.~Bekaert, and A.~A. Tseytlin, {\it {Partition function of free
  conformal higher spin theory}},  {\em JHEP} {\bf 1408} (2014) 113,
  [\href{http://arxiv.org/abs/1406.3542}{{\tt arXiv:1406.3542}}].

\bibitem{Giombi:2014yra}
S.~Giombi, I.~R. Klebanov, and A.~A. Tseytlin, {\it {Partition Functions and
  Casimir Energies in Higher Spin $AdS_{d+1}/CFT_d$}},
  \href{http://arxiv.org/abs/1402.5396}{{\tt arXiv:1402.5396}}.

\bibitem{Cappelli:1988vw}
A.~Cappelli and A.~Coste, {\it {On the Stress Tensor of Conformal Field
  Theories in Higher Dimensions}},  {\em Nucl.Phys.} {\bf B314} (1989) 707.

\bibitem{Herzog:2013ed}
C.~P. Herzog and K.-W. Huang, {\it {Stress Tensors from Trace Anomalies in
  Conformal Field Theories}},  {\em Phys.Rev.} {\bf D87} (2013) 081901,
  [\href{http://arxiv.org/abs/1301.5002}{{\tt arXiv:1301.5002}}].

\bibitem{Grisaru:1984vk}
M.~T. Grisaru, N.~Nielsen, W.~Siegel, and D.~Zanon, {\it {Energy Momentum
  Tensors, Supercurrents, (Super)traces and Quantum Equivalence}},  {\em
  Nucl.Phys.} {\bf B247} (1984) 157.

\bibitem{Fradkin:1984ai}
E.~Fradkin and A.~A. Tseytlin, {\it {Quantum Equivalence of Dual Field
  Theories}},  {\em Annals Phys.} {\bf 162} (1985) 31.

\bibitem{Ferrara:1977mv}
S.~Ferrara and B.~Zumino, {\it {Structure of Conformal Supergravity}},  {\em
  Nucl.Phys.} {\bf B134} (1978) 301.

\bibitem{Fradkin:1981jc}
E.~S. Fradkin and A.~A. Tseytlin, {\it {One Loop Beta Function in Conformal
  Supergravities}},  {\em Nucl.Phys.} {\bf B203} (1982) 157.

\bibitem{Liu:1998bu}
H.~Liu and A.~A. Tseytlin, {\it {D = 4 super Yang-Mills, D = 5 gauged
  supergravity, and D = 4 conformal supergravity}},  {\em Nucl.Phys.} {\bf
  B533} (1998) 88--108, [\href{http://arxiv.org/abs/hep-th/9804083}{{\tt
  hep-th/9804083}}].

\bibitem{Buchbinder:2012uh}
I.~Buchbinder, N.~Pletnev, and A.~Tseytlin, {\it {'Induced' $\mathcal N=4$
  conformal supergravity}},  {\em Phys.Lett.} {\bf B717} (2012) 274--279,
  [\href{http://arxiv.org/abs/1209.0416}{{\tt arXiv:1209.0416}}].

\bibitem{Henningson:1998gx}
M.~Henningson and K.~Skenderis, {\it {The Holographic Weyl anomaly}},  {\em
  JHEP} {\bf 9807} (1998) 023, [\href{http://arxiv.org/abs/hep-th/9806087}{{\tt
  hep-th/9806087}}].

\bibitem{Gunaydin:1984wc}
M.~Gunaydin, P.~van Nieuwenhuizen, and N.~Warner, {\it {General Construction of
  the Unitary Representations of Anti-de Sitter Superalgebras and the Spectrum
  of the $S^{4}$ Compactification of Eleven-dimensional Supergravity}},  {\em
  Nucl.Phys.} {\bf B255} (1985) 63.

\bibitem{vanNieuwenhuizen:1984iz}
P.~van Nieuwenhuizen, {\it {The Complete Mass Spectrum of $d=11$ Supergravity
  Compactified on $S^{4}$ and a General Mass Formula for Arbitrary Cosets
  $M^{4}$}},  {\em Class.Quant.Grav.} {\bf 2} (1985) 1.

\bibitem{Metsaev:2010kp}
R.~Metsaev, {\it {6d conformal gravity}},  {\em J.Phys.} {\bf A44} (2011)
  175402, [\href{http://arxiv.org/abs/1012.2079}{{\tt arXiv:1012.2079}}].

\bibitem{Maldacena:2011mk}
J.~Maldacena, {\it {Einstein Gravity from Conformal Gravity}},
  \href{http://arxiv.org/abs/1105.5632}{{\tt arXiv:1105.5632}}.

\bibitem{Chang:2005ska}
A.~Chang, J.~Qing, and P.~Yang, {\it {On the renormalized volumes for
  conformally compact Einstein manifolds}},
  \href{http://arxiv.org/abs/math/0512376}{{\tt math/0512376}}.

\bibitem{Nishimura:1999av}
M.~Nishimura and Y.~Tanii, {\it {Local symmetries in the AdS$_{7}$ / CFT$_{6}$
  correspondence}},  {\em Mod.Phys.Lett.} {\bf A14} (1999) 2709--2720,
  [\href{http://arxiv.org/abs/hep-th/9910192}{{\tt hep-th/9910192}}].

\bibitem{Gibbons:2006ij}
G.~Gibbons, M.~Perry, and C.~Pope, {\it {Partition functions, the Bekenstein
  bound and temperature inversion in anti-de Sitter space and its conformal
  boundary}},  {\em Phys.Rev.} {\bf D74} (2006) 084009,
  [\href{http://arxiv.org/abs/hep-th/0606186}{{\tt hep-th/0606186}}].

\bibitem{Bergshoeff:1980is}
E.~Bergshoeff, M.~de~Roo, and B.~de~Wit, {\it {Extended Conformal
  Supergravity}},  {\em Nucl.Phys.} {\bf B182} (1981) 173.

\bibitem{Coomans:2011ih}
F.~Coomans and A.~Van~Proeyen, {\it {Off-shell N=(1,0), D=6 supergravity from
  superconformal methods}},  {\em JHEP} {\bf 1102} (2011) 049,
  [\href{http://arxiv.org/abs/1101.2403}{{\tt arXiv:1101.2403}}].

\bibitem{deBoer:2009pn}
J.~de~Boer, M.~Kulaxizi, and A.~Parnachev, {\it {AdS(7)/CFT(6), Gauss-Bonnet
  Gravity, and Viscosity Bound}},  {\em JHEP} {\bf 03} (2010) 087,
  [\href{http://arxiv.org/abs/0910.5347}{{\tt arXiv:0910.5347}}].

\bibitem{Kulaxizi:2009pz}
M.~Kulaxizi and A.~Parnachev, {\it {Supersymmetry Constraints in Holographic
  Gravities}},  {\em Phys. Rev.} {\bf D82} (2010) 066001,
  [\href{http://arxiv.org/abs/0912.4244}{{\tt arXiv:0912.4244}}].

\bibitem{Romer:1985yg}
H.~Romer and P.~van Nieuwenhuizen, {\it {Axial Anomalies in $N=4$ Conformal
  Supergravity}},  {\em Phys.Lett.} {\bf B162} (1985) 290.

\bibitem{Beem:2014kka}
C.~Beem, L.~Rastelli, and B.~C. van Rees, {\it {$ \mathcal{W} $ symmetry in six
  dimensions}},  {\em JHEP} {\bf 1505} (2015) 017,
  [\href{http://arxiv.org/abs/1404.1079}{{\tt arXiv:1404.1079}}].

\bibitem{Polyakov:1981rd}
A.~M. Polyakov, {\it {Quantum Geometry of Bosonic Strings}},  {\em Phys.Lett.}
  {\bf B103} (1981) 207--210.

\bibitem{Beem:2013sza}
C.~Beem, M.~Lemos, P.~Liendo, W.~Peelaers, L.~Rastelli, et~al., {\it {Infinite
  Chiral Symmetry in Four Dimensions}},  {\em Commun.Math.Phys.} {\bf 336}
  (2015), no.~3 1359--1433, [\href{http://arxiv.org/abs/1312.5344}{{\tt
  arXiv:1312.5344}}].

\bibitem{Bergshoeff:1985mz}
E.~Bergshoeff, E.~Sezgin, and A.~Van~Proeyen, {\it {Superconformal Tensor
  Calculus and Matter Couplings in Six-dimensions}},  {\em Nucl.Phys.} {\bf
  B264} (1986) 653.

\bibitem{Romans:1986er}
L.~Romans, {\it {Selfduality for Interacting Fields: Covariant Field Equations
  for Six-dimensional Chiral Supergravities}},  {\em Nucl.Phys.} {\bf B276}
  (1986) 71.

\bibitem{Riccioni:1997np}
F.~Riccioni, {\it {Tensor multiplets in six-dimensional (2,0) supergravity}},
  {\em Phys.Lett.} {\bf B422} (1998) 126--134,
  [\href{http://arxiv.org/abs/hep-th/9712176}{{\tt hep-th/9712176}}].

\bibitem{deRoo:1984gd}
M.~de~Roo, {\it {Matter Coupling in $\mathcal N=4$ Supergravity}},  {\em
  Nucl.Phys.} {\bf B255} (1985) 515.

\bibitem{Ferrara:2012ui}
S.~Ferrara, R.~Kallosh, and A.~Van~Proeyen, {\it {Conjecture on hidden
  superconformal symmetry of $\mathcal N=4$ Supergravity}},  {\em Phys.Rev.}
  {\bf D87} (2013), no.~2 025004, [\href{http://arxiv.org/abs/1209.0418}{{\tt
  arXiv:1209.0418}}].

\bibitem{Carrasco:2013ypa}
J.~Carrasco, R.~Kallosh, R.~Roiban, and A.~Tseytlin, {\it {On the $U(1)$
  duality anomaly and the S-matrix of $\mathcal N=4$ supergravity}},  {\em
  JHEP} {\bf 1307} (2013) 029, [\href{http://arxiv.org/abs/1303.6219}{{\tt
  arXiv:1303.6219}}].

\bibitem{Beccaria:2015vaa}
M.~Beccaria and A.~Tseytlin, {\it {On higher spin partition functions}},  {\em
  J.Phys.} {\bf A48} (2015), no.~27 275401,
  [\href{http://arxiv.org/abs/1503.08143}{{\tt arXiv:1503.08143}}].

\bibitem{Erdmenger:1997gy}
J.~Erdmenger, {\it {Conformally covariant differential operators: Properties
  and applications}},  {\em Class.Quant.Grav.} {\bf 14} (1997) 2061--2084,
  [\href{http://arxiv.org/abs/hep-th/9704108}{{\tt hep-th/9704108}}].

\bibitem{Elizalde:1996nb}
E.~Elizalde, M.~Lygren, and D.~Vassilevich, {\it {Antisymmetric tensor fields
  on spheres: Functional determinants and nonlocal counterterms}},  {\em
  J.Math.Phys.} {\bf 37} (1996) 3105--3117,
  [\href{http://arxiv.org/abs/hep-th/9602113}{{\tt hep-th/9602113}}].

\bibitem{Samtleben:2012mi}
H.~Samtleben, E.~Sezgin, R.~Wimmer, and L.~Wulff, {\it {New superconformal
  models in six dimensions: Gauge group and representation structure}},  {\em
  PoS} {\bf CORFU2011} (2011) 071, [\href{http://arxiv.org/abs/1204.0542}{{\tt
  arXiv:1204.0542}}].

\bibitem{Fradkin:1982kf}
E.~Fradkin and A.~A. Tseytlin, {\it {Quantum Properties of Higher Dimensional
  and Dimensionally Reduced Supersymmetric Theories}},  {\em Nucl.Phys.} {\bf
  B227} (1983) 252.

\bibitem{Graham:2007}
C.~Graham, {\it {Conformal powers of the Laplacian via stereographic
  projection}},  {\em SIGMA} {\bf 3} (2007) 066,
  [\href{http://arxiv.org/abs/0711.4798}{{\tt arXiv:0711.4798}}].

\bibitem{Manvelyan:2006bk}
R.~Manvelyan and D.~Tchrakian, {\it {Conformal coupling of the scalar field
  with gravity in higher dimensions and invariant powers of the Laplacian}},
  {\em Phys.Lett.} {\bf B644} (2007) 370--374,
  [\href{http://arxiv.org/abs/hep-th/0611077}{{\tt hep-th/0611077}}].

\bibitem{Juhl:2011aa}
A.~{Juhl}, {\it {Explicit formulas for GJMS-operators and \$Q\$-curvatures}},
  {\em ArXiv e-prints} (Aug., 2011) [\href{http://arxiv.org/abs/1108.0273}{{\tt
  arXiv:1108.0273}}].

\bibitem{Pang:2012rd}
Y.~Pang, {\it {One-Loop Divergences in 6D Conformal Gravity}},  {\em Phys.Rev.}
  {\bf D86} (2012) 084039, [\href{http://arxiv.org/abs/1208.0877}{{\tt
  arXiv:1208.0877}}].

\end{thebibliography}\endgroup
\bibliographystyle{JHEP}

\end{document}

\iffa 
\be 
(6; 2,0,0) \longrightarrow
 \det\hat\Delta_{2\perp}(\{8,6,2\})\,
 \det\hat\Delta_{1\perp}(\{7,5,1\})\,
 \det\hat\Delta_{0}(\{6,4,0\}).
\ee
The relevant ghost representation is, again from (\ref{A.11}),
\be
(7; 1,0,0) \longrightarrow
 \det\hat\Delta_{1\perp}(\{7,5,1,-5\})\,
 \det\hat\Delta_{0}(\{6,4,0,-6\}).
\ee
Thus, for the conformal graviton in 6d
\fi

\iffa 
\section{Non-unitary conformal symmetric spinor-tensor theories in 4d}

As we mentioned in App.~\ref{app:s1d6}, the result of \cite{Erdmenger:1997wy}
is a curved space action for  a rank $s$ conformal 
symmetric tensor (CST) $\varphi_{s}\equiv \varphi_{\mu_{1}, \dots, \mu_{s}}$
with a 2-derivatives kinetic term. 
If the background is flat and $d=4$, this action admits a minimal amount of (scalar) gauge invariance.
The analysis of conformal anomalies and AdS/CFT interpretation of the CST theories has been recently
presented in \cite{Beccaria:2015vaa}.
In this Appendix, we discuss their fermionic generalization where the fundamental field is a
conformal symmetric spinor-tensor (CSST) $\psi_{s}\equiv \psi_{\mu_{1}, \dots, \mu_{s}}$.
To begin, let us remind some more details of what happens in the bosonic case. 
The  action computed in \cite{Erdmenger:1997wy} has the following properties: 
\begin{enumerate}
\item minimal kinetic term $S = \int d^{d}x\,\sqrt{g}\,\varphi_{s} \nabla^{2}\, \varphi_{s}+\cdots$,
\item scalar gauge invariance on conformally flat background $\delta\varphi_s = \partial^{s}\sigma$,
\item local Weyl invariance: $g'_{\mu\nu}=\Omega^{2}g_{\mu\nu}$, $\varphi'_{s} = \Omega^{s-\frac{1}{2}(d-2)}
\varphi_{s}$. \footnote{
For an action $\sim \int d^{d}x \sqrt{g}\varphi_{s}(\nabla^{2})^{n}\varphi_{s}$, we have $\sqrt{g}\to \Omega^{d}\sqrt{g}$. Also each $\nabla^{2}\to \Omega^{-2}\nabla^{2}$ because of the hidden $g^{\mu\nu}$. Similarly, we have 
another factor $\Omega^{-2}$ for each index in $\varphi_{s}$. Thus we have invariance under $\varphi_{s}\to \Omega^{\gamma}\varphi_{s}$ if $d-2n-2s+2\gamma=0$ or $\gamma = s+n-\frac{d}{2}$.}
\end{enumerate}
This construction is certainly meaningful for $s\ge 1$. One can try to extend it to $s=0$ where it is  possible
to satisfy (1) and (3). However, for spin zero, (2) is meaningless. For this reason, we shall assume the construction to be 
defined for $s\ge 1$. By analogy we shall replace (1-3) in fermionic case by the following properties:
\begin{enumerate}
\item[1'.] minimal kinetic term $S = \int d^{d}x\,e\,\overline\psi_{s} \nabla^{2}\slashed{\nabla}\, \psi_{s}+\cdots$,
where $\psi_{s}$ has $\s=s-\frac{1}{2}$ symmetric  vector indices, and is $\gamma$-traceless,
\item[2'.] spin $\frac{1}{2}$ gauge invariance on conformally flat background $\delta\psi_s = \partial^{\s}\chi$,
\item[3'.] local Weyl invariance: $g'_{\mu\nu}=\Omega^{2}g_{\mu\nu}$, $\psi'_{s} = \Omega^{s-\frac{1}{2}(d-2)}
\psi_{s}$.
 \footnote{
For a fermionic action $\sim \int d^{d}x \sqrt{g}\overline\psi_{s}(\nabla^{2})^{n}\slashed{\nabla}\psi_{s}$ where $\psi_{s}$ has $\s=s-\frac{1}{2}$ indices, we need $d-2n-2\s-1+2\gamma=0$ or $\gamma = n+\s+\frac{1}{2}-\frac{d}{2} = n+s-\frac{d}{2}$ which is the same as in bosonic case.}
\end{enumerate}
The case $s=\frac{3}{2}$ is conformal gravitino. For half-integer $s>\frac{3}{2}$, we shall assume without proof that a unique generalisation of  the action in \cite{Erdmenger:1997wy} does exist obeying (1'-3'). 
Under this assumption, we shall compute the conformal anomalies and Casimir energy by exploiting the 
connection with HS fields on AdS$_{5}$ according to the approach recalled in Sec.~\ref{sec:2}.

From a representation-theoretical point of view, we are considering conformal fermionic fields
of different  {\em depth} \cite{Bekaert:2013zya,Beccaria:2015vaa}. In this language, conformal gravitino field is
minimal depth, while the Deser-Nepomechie gravitino \cite{Deser:1983tm}, to be discussed in \ref{sec:B2}, is maximal depth.

%
%
%
\subsection{CSST conformal anomalies}

\red{do we need to recall the 4d definition of a and c anomalies ? }

We recall the general formula for the a-anomaly of a 4d conformal field whose HS counterpart in AdS$_{5}$
transforms in the $SO(4,2)$ representation $(\Delta; j_{1}, j_{2})$  
\cite{Beccaria:2014xda}
\be
\label{B.1}
\begin{split}
{\aa}(\Delta; j_{1}, j_{2}) &= \frac{(-1)^{2(j_{1}+j_{2})}}{720}(2j_{1}+1)(2j_{2}+1)(\Delta-2)
\Big[-3(\Delta-2)^{4}\\
&+10(j_{1}^{2}+j_{2}^{2}+j_{1}+j_{2}+\frac{1}{2})(\Delta-2)^{2}
-15(j_{1}-j_{2})^{2}(j_{1}+j_{2}+1)^{2}\Big].
\end{split}
\ee
The fermion $\psi_{s}$ of the CSST theory has dimension $\frac{1}{2}$. Thus, the associated HS field 
transforms in the representation 
\be
\label{B.2}
\Big(\frac{7}{2}; \frac{\s+1}{2}, \frac{\s}{2}\Big)+\Big(\frac{7}{2}; \frac{\s}{2}, \frac{\s+1}{2}\Big).
\ee
The gauge invariance $\delta\psi_{s} = \partial^{\s}\chi$ is  associated to the spin $\frac{1}{2}$
representation
\be
\label{B.3}
\Big(\frac{7}{2}+\s; \frac{1}{2}, 0\Big)+\Big(\frac{7}{2}+\s; 0, \frac{1}{2}\Big).
\ee
Subtracting the two values of $\aa$ corresponding to (\ref{B.2}) and (\ref{B.3}), we obtain 
\be
\label{B.4}
\aa_{s}^{\text{CSST}} = -\frac{\s}{1440}\,(24\,\s^{4}+225\,\s^{3}+710\,\s^{2}+882\,\s+351).
\ee
Taking $s=\frac{3}{2}$ or $\s=1$, we obtain $\aa_{\frac{3}{2}} = -\frac{137}{90}$, correct for the 4d conformal gravitino.
Now, we want to re-interpret (\ref{B.4}) as the anomaly associated to a specific factorisation of the 4d Lagrangian 
on $S^{4}$. The building blocks are the operators defined in \cite{Tseytlin:2013jya}. In particular, we need the mass terms
\be
\label{B.5}
m^{2}_{\s, \kk} = \s+3-(\kk+1)^{2},
\ee
and the anomaly associated to the Laplace operator, with mass term, acting on $\gamma$-transverse traceless
spinor-tensors with spin $s=\s+\frac{1}{2}$
\be
\label{B.6}
\aa[\hat\Delta_{\s\perp}(M^{2})] = -\frac{1}{720}\,(\s+1)(-101+20\,\s\,(3\,\s^{2}+13\,\s+11) 
-60\,(\s^{2}+\s-2)\,M^{2}-30\,M^{4}).
\ee
Indeed, it is possible to write (\ref{B.4}) in the form 
\be
\label{B.7}
\aa_{s}^{\text{CSST}} = \sum_{\kk=1}^{\s}\Big[
2\,\aa[\hat\Delta_{\kk\perp}(m^{2}_{\kk, 0})]+\aa[\hat\Delta_{\kk\perp}(m^{2}_{\kk, -1})]
-2\,\aa[\hat\Delta_{0}(m^{2}_{0,\kk})].
\Big]
\ee
The interpretation of (\ref{B.7}) is that the first two terms take into account the three derivatives operator acting on 
all the spin $s$, $s-\frac{1}{2}$, $\dots$ components of $\psi_{s}$, while the last term is the subtraction of the spin $\frac{1}{2}$ ghost field. With the notation of \cite{Tseytlin:2013jya}, this means the following partition function
\be
\label{B.8}
Z_{s} = \prod_{k=\frac{3}{2}}^{s}\left[
\frac{\det^{2}\hat\Delta_{\frac{1}{2}}(m^{2}_{\frac{1}{2}, k})}
{\det^{2}\hat\Delta_{k\perp}(m^{2}_{k,\frac{1}{2}})\,\det\hat\Delta_{k\perp}(m^{2}_{k,\emptyset})}
\right]^{-1/4}.
\ee
Let us recall that for a CST field with integer spin $s$ we have the same count of dynamical (on-shell)
degrees of freedom as for CHS (in 4d)
\be
\la{B.9}
\nu^{(\rm b)}_{\rm CHS} = \nu^{(\rm b)}_{\rm CST} = s(s+1).
\ee
For a  CSST field,  the number of off-shell degrees of freedom is 
\be
\la{B.10}
N_{\s\perp} = 4\,(\s+1),
\ee
and, using (\ref{B.8}), we find
\be
\la{B.11}
\nu^{(\rm f)}_{\rm CSST} = \frac{1}{2}\,\sum_{\kk=1}^{\s}(3\,N_{\kk\perp}-8) = \s\,(3\,\s+5) = \Big(s-\frac{1}{2}\Big)\,
\Big(3s+\frac{7}{2}\Big).
\ee
The c-anomaly can also be computed from the general expressions derived in \cite{Beccaria:2014xda} applied to the subtraction 
(\ref{B.2})-(\ref{B.3}). This gives
\be
\la{B.12}
\cc_{{\rm CSST}, s} = -\frac{1}{11520}\, (2 s+1)\, (64 s^6+464 s^5+832 s^4+104 s^3-692 s^2-247 s+264).
\ee
This expressions must be taken with some caution due to possible ambiguities discussed at length in 
\cite{Beccaria:2014xda,Beccaria:2015vaa}.

\subsection{The Deser-Nepomechie conformal gravitino action}
\label{sec:B2}

We have presented a full discussion of actions with kinetic term of the form 
$\psi_{s} \nabla^{2}\slashed{\nabla}\, \psi_{s}$. This is the minimal possibility if we want to 
have minimal (spin $\frac{1}{2}$) gauge invariance. For a completely non gauge invariant action, 
it is possible to further lower the order of derivatives. As an interesting example, we work here the case 
of the gravitino action discussed in Eq.~(10) of \cite{Deser:1983tm}.
It reads
\be
\label{B.13}
S_{\text{DN}, \frac{3}{2}} = -\frac{i}{2}\int d^{4}x\,e\,\overline{\widetilde\psi}^{\mu} \,\slashed{\nabla}\,\widetilde\psi_{\mu},
\ee
where $\widetilde\psi_{\mu}=\psi_{\mu}-\frac{1}{4}\gamma_{\mu}\,\gamma\cdot\psi$ is $\gamma$-traceless.
This action is Weyl invariant, but not gauge invariant. \footnote{We remind again that the conformal gravitino
has instead a kinetic term $\sim\overline\psi^{\mu}\nabla^{2}\slashed{\nabla}\psi_{\mu}$.}
Thus, taking into account the different dimension of $\psi_{\mu}$ compared with conformal gravitino, we find that the a-anomaly is given by 
\be
\label{B.14}
\aa_{{\rm DN}, \frac{3}{2}} = 2\,\widehat{\aa}(\tfrac{5}{2}, 1, \tfrac{1}{2}) = \frac{31}{240}.
\ee
To recover this expression from the explicit factorization of the action (\ref{B.13}) on $S^{4}$, one must split
the gravitino field into a transverse part plus a $\gamma$-traceless component:
\be
\label{B.15}
\psi_{\mu} = \psi_{\mu\perp}+(\nabla_{\mu}-\frac{1}{4}\gamma_{\mu}\,\slashed{\nabla})\,\chi,
\ee
and substitute in the action (\ref{B.13}). The field $\psi_{\mu\perp}$ gives a contribution involving
$\Delta_{\frac{3}{2}\perp}(4)$. The spin $\frac{1}{2}$ field $\chi$ is decoupled.
%
%
After a short calculation, the partition function can be written
\be
\la{B.16}
Z_{\text{DN}, \frac{3}{2}} = \Big[
\frac{(\det\hat\Delta_{\frac{1}{2}}(-1))^{2}}{\det\hat\Delta_{\frac{3}{2}\perp}(4)\,(\det\hat\Delta_{\frac{1}{2}}(-1))^{2}\,
\det\hat\Delta_{\frac{1}{2}}(3)]}
\Big]^{-1/4} = \Big[
\frac{1}{\det\hat\Delta_{\frac{3}{2}\perp}(4)\,
\det\hat\Delta_{\frac{1}{2}}(3)]}
\Big]^{-1/4},
\ee
where the numerator in the middle expressions comes from the Jacobian for the splitting (\ref{B.15}).
Thus, from  (\ref{B.6}), we  obtain \footnote{
We remind that for the conformal gravitino, we had 
\be
\notag
\aa_{\frac{3}{2}} = \aa[\hat\Delta_{\frac{3}{2}\perp}(4)]+2\,\aa[\hat\Delta_{\frac{3}{2}\perp}(3)]-2\,
\aa[\hat\Delta_{\frac{1}{2}}(-1)] = -\frac{137}{90}.
\ee
Here, the first two terms come from the factorization of the $\slashed{\nabla}^{3}$ operator, while the last term is 
a Jacobian for the splitting (\ref{B.15}).
}
\be
\label{B.17}
\aa_{{\rm DN},\frac{3}{2}} = \aa[\hat\Delta_{\frac{3}{2}\perp}(4)]
+\aa[\hat\Delta_{\frac{1}{2}}(3)] = 
\aa[\hat\Delta_{\frac{3}{2}\perp}(m^{2}_{\frac{3}{2}, \emptyset})]
+\aa[\hat\Delta_{\frac{1}{2}}(m^{2}_{\frac{1}{2}, \emptyset})] =  \frac{31}{240},
\ee
where we expressed the mass terms by  the characteristic fermionic masses with half-integer indices (see the expression (\ref{B.5}) in terms of shifted indices) 
\be
\la{B.18}
m^{2}_{s,k} = s+\tfrac{5}{2}-(k+\tfrac{1}{2})^{2}, \qquad m^{2}_{s,\emptyset} = m^{2}_{s,-\tfrac{1}{2}} = 
s+\tfrac{5}{2}.
\ee
The fact that there are no {\em subtractions} in (\ref{B.17}) may be explained by noting that there is no gauge invariance here. Usually, subtraction terms can be associated with covariant ghost terms, but here they are simply absent.

\bigskip
\noindent
The c-anomaly also follows from the factorization (\ref{B.16}), {\em i.e.} from the two terms in (\ref{B.17}).
Using the formulas for c derived in \cite{Beccaria:2014xda} (see Eq. 4.9 therein), one obtains 
\be
\label{B.19}
\cc_{{\rm DN}, \frac{3}{2}} = -\frac{7}{40}.
\ee
This value is free from the ambiguities discussed in  \cite{Beccaria:2014xda,Beccaria:2015vaa}
due to the low value of the spin.
It can be obtained quite easily starting from the
factorization  (\ref{B.16}) and going to a Ricci flat background. We can use 
\be
\la{B.20}
\det\hat\Delta_{\text{L},s\perp} = \frac{\det\hat\Delta_{\text{L},s}}{\det\hat\Delta_{\text{L},s-1}},
\ee
where $\hat\Delta_{\text{L}, s}$ is the natural counterpart of spin 2 Lichnerowicz operator on Ricci flat background 
\cite{Christensen:1978md,Tseytlin:2013jya}.
We see that there is full cancellation of the spin $\frac{1}{2}$ piece. Finally, using  $\cc-\aa=\beta_{1}$ where,
for fermions, we have \cite{Tseytlin:2013jya} 
\be
\la{B.21}
\beta_{1, s} = \frac{1}{2880}N_{s}(50-28N_{s}+3N_{s}^{2}), \qquad N_{s} = 2\,(s+\tfrac{1}{2})(s+\tfrac{3}{2}),
\ee
we recover  \footnote{Notice that $\beta_{1,s}$ computed in  \cite{Tseytlin:2013jya}  is associated with 
$(\det\hat\Delta_{\text{L},s})^{1/2}$, while here we have $(\det\hat\Delta_{\text{L}, \frac{3}{2}})^{-1/4}$.}
\be
\la{B.22}
\cc_{{\rm DN}, \frac{3}{2}} = \aa_{{\rm DN}, \frac{3}{2}}-\frac{1}{2}\,\beta_{1}(1,\tfrac{1}{2}) = \frac{31}{240}-
\frac{73}{240} = -\frac{7}{40},
\ee
in agreement with (\ref{B.19}).

\fi 

%
%
%
%
%